\documentclass[aps,prb,amsmath,amssymb,twocolumn,letterpaper,
         footinbib,bibnotes,showpacs,reprint,balancelastpage,raggedbottom,longbibliography,%
         citeautoscript,floatfix]{revtex4-1}
\usepackage[usenames,dvipsnames,table]{xcolor}
\usepackage[%
  breaklinks,             %
  bookmarks = false, %
  pdfpagemode   = UseNone,%
  pdfstartview  = FitH,     %
  pdfstartpage  = 1,      %
  colorlinks    = true,   %
]{hyperref}
\hypersetup{
    linkcolor=RubineRed,          
    citecolor=RoyalBlue,        
    filecolor=Mulberry,      
    urlcolor=RoyalBlue           
}

\usepackage[protrusion=true,expansion=true,final]{microtype}
\usepackage{amssymb,amsmath,bm}
\usepackage{graphicx}
\usepackage{booktabs}
\usepackage{float}
\usepackage{enumerate}
\usepackage{microtype}
\usepackage{xspace}
\usepackage{nicefrac}
\usepackage{soul}  
\usepackage[utf8]{inputenc}
\usepackage{pdfpages,pgffor}
\DeclareUnicodeCharacter{2212}{-}

\makeatletter

\AtBeginDocument{\let\LS@rot\@undefined} 
\makeatother

\begin{document}
\author{Yongjin Shin}
\affiliation{Pritzker School of Molecular Engineering, University of Chicago, Chicago, Illinois 60637, USA}
\email{yongjinshin@uchicago.edu}
 \author{Kenneth R. Poeppelmeier}
 \affiliation{Department of Chemistry, Northwestern University, Evanston, Illinois 60208, USA}
\author{James M.\ Rondinelli}
 \affiliation{Department of Materials Science and Engineering, Northwestern University, Evanston, Illinois 60208, USA}
 \email{jrondinelli@northwestern.edu}

\def\degree{$^\circ$\xspace}
\newcommand\Tstrut{\rule{0pt}{2.6ex}}         
\newcommand\Bstrut{\rule[-0.9ex]{0pt}{0pt}}   
\def\CAO{CaAlO$_{2.5}$\xspace}
\def\SFO{SrFeO$_{2.5}$\xspace}
\def\SMO{SrMnO$_{2.5}$\xspace}
\def\ABO{$AB$O$_{2.5}$\xspace}
\def\PC{$_\mathrm{pc}$\xspace}

\newcommand{\supf}{\textcolor{red}{Figure S}}
\newcommand{\supt}{\textcolor{red}{Table S}}

%

\title{Informatics-based learning of oxygen vacancy ordering principles in oxygen-deficient perovskites}
%


\begin{abstract}
Ordered oxygen vacancies (OOVs) in perovskites can exhibit long-range order and may be used to direct materials properties through modifications in electronic structures and broken symmetries.
Based on the various vacancy patterns observed in previously known compounds, we explore the ordering principles of OOVs in oxygen-deficient perovskite oxides with \ABO stoichiometry  to identify other OOV variants. We performed first-principles calculations to assess the OOV stability on a dataset of 50 OOV structures generated from our bespoke algorithm.
The algorithm employs uniform planar vacancy patterns on (111) pseudocubic perovskite layers and the approach proves effective for generating stable OOV patterns with minimal computational loads.
We find as expected that the major factors determining the stability of OOV structures include coordination preferences of transition metals and elastic penalties resulting from the assemblies of polyhedra. 
Cooperative rotational modes of polyhedra within OOV structures reduce elastic instabilities by optimizing the bond valence of $A$- and $B$-cations.
This finding explains the observed formation of vacancy channels along low-index crystallographic directions in prototypical OOV phases.
The identified ordering principles enable us to devise other stable vacancy patterns with longer periodicity for targeted property design in yet to be synthesized compounds. 
\end{abstract}

\maketitle

\section{Introduction}

The perovskite structure has served as a rich base for novel materials phenomena and relevant applications in transition metal oxides because of its versatile electronic, magnetic, optical, and ionic properties \cite{Goodenough1963,Ahn2021,Ramanathan2022, Rondinelli2011,Tokura2008,Forst_nonlinear_2011}.
The origin of versatile tunability of complex oxides lies in localized electrons within the $d$-orbitals of transition metals, whose energy states and orbital overlaps change sensitively in response to variations in bond characteristics \cite{Li/Hwang2019,Chakhalian2012,Zubko2011}.
The sensitive interplay between atomic structure and properties enables us to achieve the desired properties by tuning atomic bond characteristics via introduction of heterointerfaces or chemical substitutions \cite{Hwang2012,Kim/Eom2016}.
In addition to cation substitution and chemical ordering to fine-tune such properties,
an alternative strategy is
utilizing anion engineering to directly modify the bond characteristic of transition metals through anion substitution 
\cite{Kageyama2018,Wang/Shin2021, Szymanski/Walters/Rondinelli2019,Laurita/Puggioni/Rondinelli:2019,Harada/Rondinelli:2019a,Charles/Saballos/Rondinelli:2018}.

\begin{figure}[t]
\centering
\includegraphics[width=0.85\columnwidth]{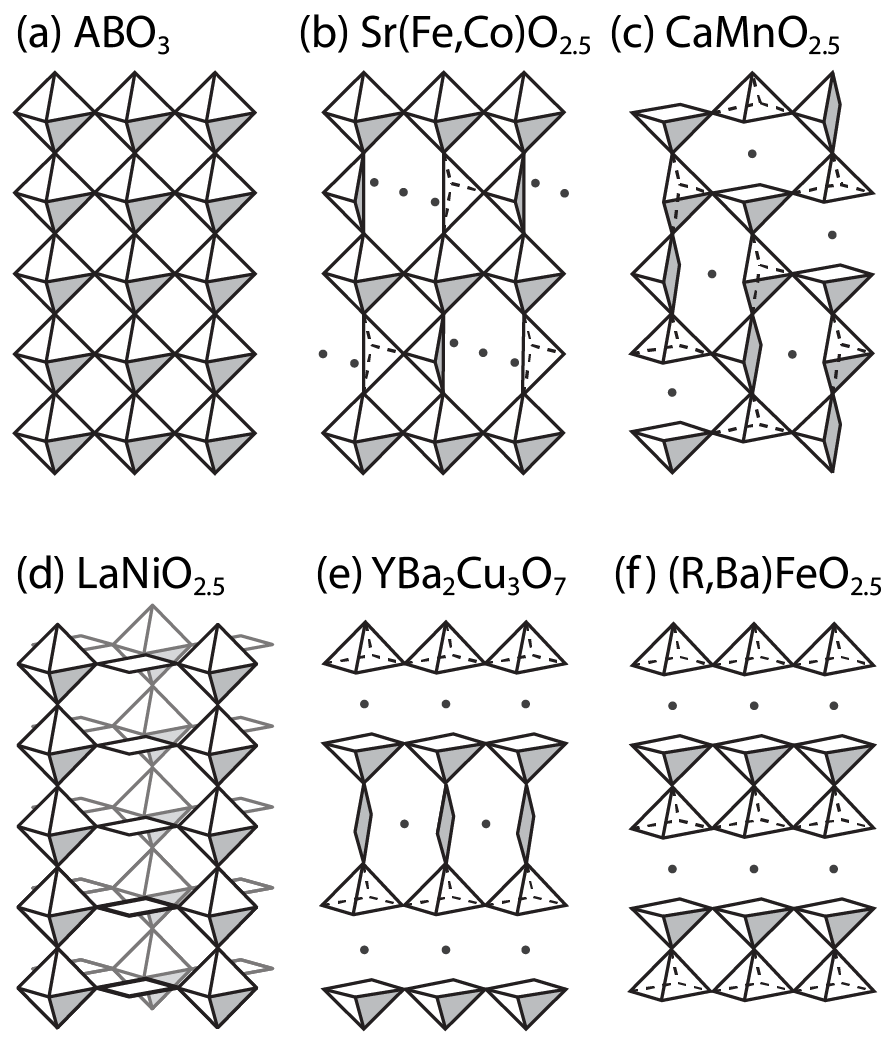}
\caption{Schematic illustration of (a) $AB$O$_3$ perovskites structures and its derivative OOV phases:
(b) Sr(Fe,Co)O$_{2.5}$ (brownmillerite), (c) Ca$_2$Mn$_2$O$_5$-type, (d) LaNiO$_{2.5}$-type, (e) YBa$_2$Cu$_3$O$_7$, and (f) ($R$,Ba)FeO$_{2.5}$ ($R$=Ho, Nd) structures based on polyhedral connectivities. $A$-cations are omitted for visibility and vacancy sites are drawn as gray dots.
}
\label{fig:schematic}
\end{figure}

Extreme tuning of bond characteristics can be realized by completely removing some of the oxygen ligands.
Although random oxygen vacancy formation occurs in the form of thermodynamic point defects, which modify the local structure and some average properties in materials \cite{Aschauer2013,Park/Chadi1998,Li/Chen2022,Wexler/Carter2021},
long-range ordered vacancies at high concentrations lead to new structure-types and well-defined compounds\cite{Anderson1993}, i.e., the vacancy acts as another anionic species and  can be exploited for electronic structure design  \cite{Kageyama2020,Stolen2006,Yamamoto2012,Shin/Rondinelli2022a,Wang/Shin2021,Lu/Yildiz2020}.
The high concentration of ordered oxygen vacancies (OOVs) is realized by the flexible charge state of transition metals, and the OOV concentration is largely limited by charge neutrality. 
When OOVs are formed in perovskites, the removal of an oxygen atom from the O$^{2-}$ site results in reduction of nearby transition metals and reconstruction of the crystal field splitting of $d$-orbital defined by the various types of functional building units (FBUs), including octahedra, square pyramids, tetrahedra, and square planes\cite{Anderson1993}.
Prototypical OOV structures are shown in \autoref{fig:schematic}, including SrFeO$_{2.5}$, SrCoO$_{2.5}$, CaMnO$_{2.5}$, LaNiO$_{2.5}$, and YBa$_2$Cu$_3$O$_7$.

Various phenomena and promising performances of OOV phases have been reported, including dual-ion conduction, ferroelectricity, magnetic transitions, insulator-metal transitions. \cite{Lu2017,PuYu2022,Young2017,Wang/Shin2018,Wang/Shin2019,Shin/Rondinelli2022a,Leighton_magnetoresistance2020,Shin/Galli2023}
Furthermore, the recent discovery of superconductivity in the infinite-layer nickelates with $R$NiO$_2$-structure \cite{Wang/Hwang2021, Li/Hwang2019, Goodge_resolving_2023,Lee/HwangNickelates2023,Hepting2020,Lu/Lee2021} highlights the promising potential of OOV structures to host novel materials phenomena.
A noteworthy application of OOV phases are for memristor materials in neuromorphic computing, taking advantage of the dramatic change in resistivity when OOVs are formed in perovskites.\cite{Zhang/Galli2020,Yao/Dijken2017,Nallagatla/Dittman2019,Nallagatla/Chang2020}
Because OOV phases are formed `topotactically' without disrupting the perovskite framework,\cite{Jeen/Lee2013,Lu/Yildiz2016}
pairing an oxide perovskite with a desired OOV phase can realize built-in memory function within materials.\cite{Han_control_2022,May/Leightron2022,Chaturvedi/Leighton2021,Yao/Dijken2017}
Given the rich opportunities with OOV phases, the discovery of various alternative OOVs vacancy patterns could further expand the range of properties in oxide perovskites.

Despite the promising aspects of OOV phases, a fundamental question that persists concerns identifying the factors governing the stability of specific vacancy patterns.
Anderson \textit{et al.}\cite{Anderson1993} have observed the structural similarities among the known OOV phases that each OOV phase exhibits specific layer vacancy pattern on the close-packed $A$O$_3$ layer.
They also identified the preferred stacking sequence of oxygen-deficient layers depends on the coordination number, cation size, and electronic configurations of $A$ and $B$ cations,
which displaying the sensitive nature of the stability of OOV patterns.
In addition, some transition metals are also known to stabilize various OOVs, extending beyond the `simple' vacancy patterns appearing in \autoref{fig:schematic},
which suggests the possibility of discovering new types of OOV phases.\cite{Aasland1998,Genouel/Raveau1995,Kwon/Jeong2020,ShaoHorn2013,Grenier2,Inoue/Kageyama2008,Yamamoto2012,Reller/Poeppelmeier1982}
Indeed, the vacancy order in \autoref{fig:schematic}(f) can be obtained in  ferrites, manganites, and cobaltites with appropriate cation mixing \cite{Woodward2003,Sengodan/Choi2015,Choi2013}.
Given the various types of OOVs,\cite{Anderson1993,Stolen2006,Yamamoto2012}
it is crucial to establish their ordering principles.

In this work, we present a comprehensive exploration of ordering principles of oxygen vacancies in OOV structures with $AB$O$_{2.5}$ chemistry.
First, we formulate and implement an efficient structure generation algorithm informed from an experimental observations \cite{Anderson1993}, which generates various assemblies of polyhedral units with minimal computational costs.
Then, we proceed by uncovering the significant factors contributing to the stability of OOV patterns using density functional theory (DFT) calculations on the generated database of structures.
The main factors are the coordination environment of the transition metal and the formation of vacancy channels, which reduce elastic instabilities in the OOV structures. 
Lastly we apply the ordering principles to generate new OOV patterns and provide an illustrative example of selecting appropriate chemical identities to stablize the targeted OOV structure.
Our data-driven study provides useful materials chemistry insights for the design of oxygen-deficient perovskites, accelerating the discovery of functional materials.

\begin{figure*}[t]
\centering
\includegraphics[width=0.75\paperwidth,clip]{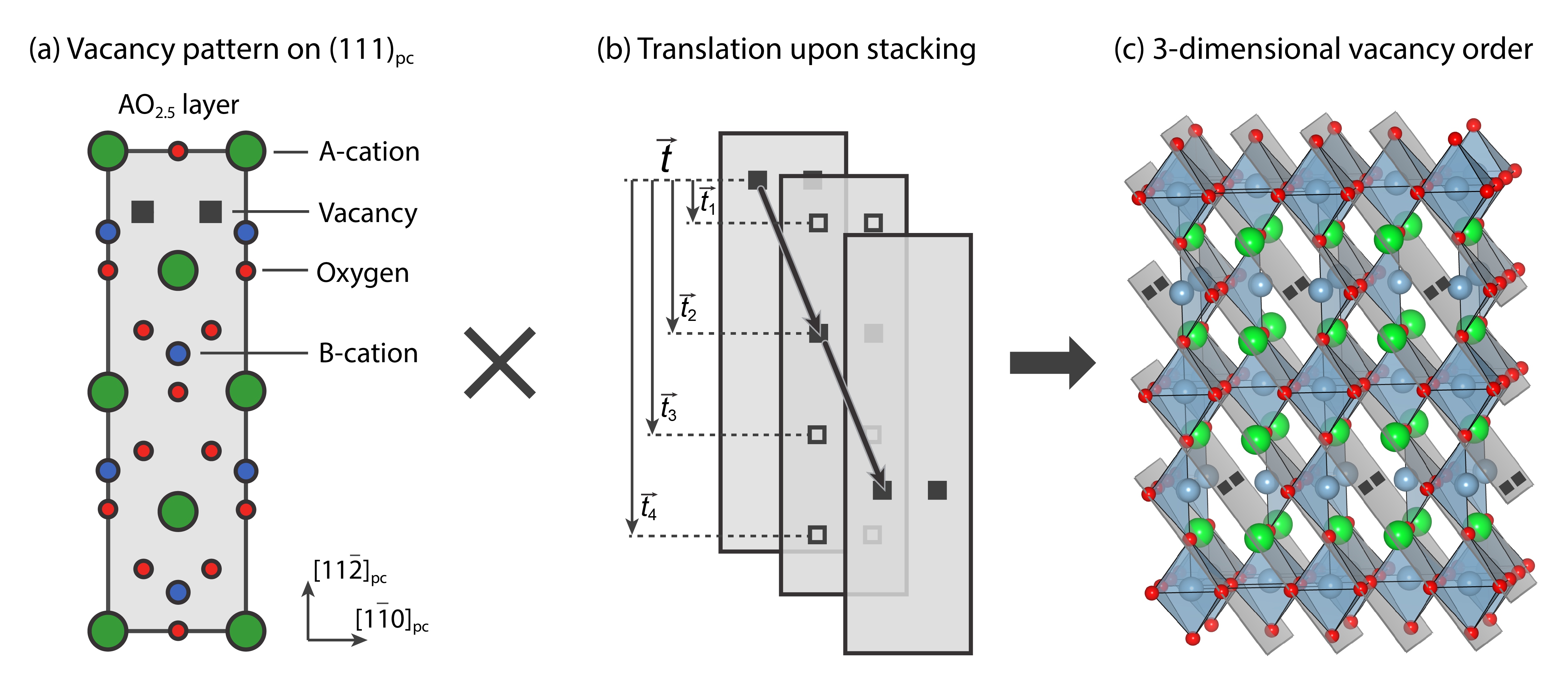}
\caption{Schematic illustration of the structure generation algorithm. (a) A planar vacancy pattern of $A$O$_{2.5}$ on the (111)$_\mathrm{pc}$ plane is defined initially. Note that in this projection the $B$-cations are on different layers from the $A$ cation and oxygen atoms; they are included to better describe the connection to the 3D structure. (b) (111)$_\mathrm{pc}$ layers are stacked to reconstruct the close-packed lattice of perovskites, where layer vacancy patterns are translated by one of many possible translation vectors ($\vec{t}$) upon stacking.  Other possible translation vectors are depicted with empty squares, indicating varying vacancy positions with different $\vec{t}$. (c) The combination of the planar vacancy pattern (2D) and stacking of these planes with the translational vector (1D) realizes a structure with 3D OOVs. The $A$ cation in the resultant structure is omitted for visibility of polyhedral units in the OOV structure.
}
\label{fig:Anderson_scheme}
\end{figure*}

\section{Computational Methods}

\subsection{Structure generation algorithm}
We construct $AB$O$_{2.5}$ structures 
by creating a three-dimensional (3D) vacancy arrangement by combining two-dimensional (2D) patterns on (111)$_\mathrm{pc}$ layers with a translation vector upon layer stacking, as shown in \autoref{fig:Anderson_scheme}. The area of the (111)$_\mathrm{pc}$ plane is prepared with size $\sqrt{2}a\times2\sqrt{6}a$, where $a$ denotes the lattice parameter of a cubic perovskite, and is initially set to 4\,\AA\xspace.
We then identified 9 unique 2D planar vacancy patterns observed in $A$O$_{2.5}$ compounds, as provided in the Supporting Information (SI) \supf{1}. 
With this specified size of the 2D plane, there are four available translation vectors, $\vec{t}$ in \autoref{fig:Anderson_scheme}(b), when stacking (111)$_\mathrm{pc}$ layers to construct a perovskite-like structure.
In addition, there are four $A$O$_{2.5}$ formula units in the unit plane, so there are four translational vectors available, $\vec{t_1}$ to $\vec{t_4}$ in \autoref{fig:Anderson_scheme}(b), to stack the planes and maintain the cation-anion ratio. 
Out of the nine patterns we identified, two of them have shorter periodicities and can be reduced to a $\sqrt{2}a\times\sqrt{6}a$ unit plane size, where only two translational vectors are available ($\vec{t_1}$ and $\vec{t_2}$).
We also note that some planar patterns exhibit handedness.
The right- and left-handed planar patterns are equivalent vacancy orderings related by a  vertical mirror operation on the $A$O$_{2.5}$ unit plane.
When stacking such arrangements, we additionally considered stacking the layer and alternating the handedness.
Compiling all of these considerations, we identified 50 ideal OOV structures with OOVs on (111)$_\mathrm{pc}$ planes.
The 50 generated structures are listed in \supt{1}, 
and provided as crystallographic information files (CIFs) on 
Github \cite{Shin/OOV_StrGen}.

\subsection{First-principles calculations}
After obtaining the ideal OOV structure set, we performed density functional theory (DFT) calculations as implemented in the Vienna Ab-initio Simulations Package (\textsc{vasp}) \cite{Kresse1996VASP,Kresse1999} with the Perdew-Burke-Ernzerhof functional (PBE)\cite{PBE1996}.
We relax the atomic structures of all 50 OOV geometries with CaAlO$_{2.5}$ and SrFeO$_{2.5}$ compositions, and relaxed a subset of the 50 geometries for LaNiO$_{2.5}$ and SrMnO$_{2.5}$ compositions. 
We used projector-augmented wave (PAW) potentials \cite{Blochl1994} to describe the electron core-valence interactions with the following configurations: 
Ca (3s$^2$ 3p$^6$ 4s$^2$), 
Sr ($4s^2 4p^6 5s^2$),  
La ($4f^0 5s^2 5p^6 5d^1 6s^2$),
Mn ($3d^6 4s^1$), 
Fe ($3d^7 4s^1$), 
Ni ($3d^9 4s^1$), 
and O ($2s^2 2p^4$).
A 550\,eV planewave cutoff was used to obtain the ground structures for both relaxation and self-consistent total energy calculations.
Brillouin zone integrations employed the tetrahedron method \cite{Bloch1994Tetrahedron}, where  $k$-point meshes were generated based using Monkhorst-Pack formalism and the $K$-point grid server with 34\,\AA\, of minimum distance between points \cite{Monkhorst1976,Mueller2016}.
The cell volume and atomic positions were evolved until the forces on each atom were less than 0.1\,kbar and 3\,meV\,\AA$^{-1}$, respectively, while maintaining the initial symmetry of the generated OOV hettotypes. 
For compounds with partially filled $d$-orbitals, we applied a Hubbard $U$ correction following the Dudarev method \cite{Dudarev1998} with 5\,eV for Fe, 3\,eV for Mn, and 1.5\,eV for Ni on the $d$ orbitals to better describe electron-electron interactions.

\begin{figure}[h]
\centering
\includegraphics[width=0.34\paperwidth,clip]{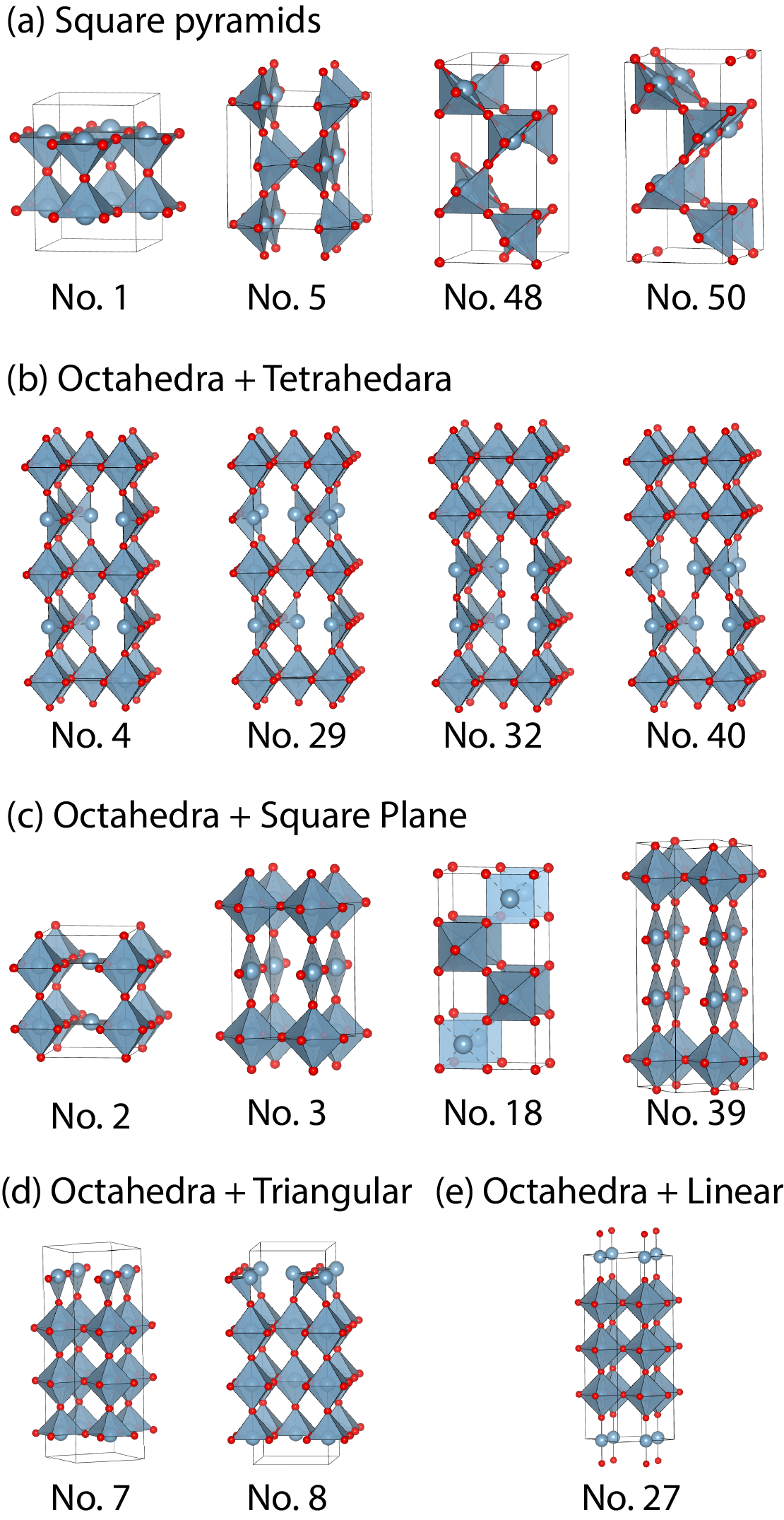}
\caption{Examples structures sampled from the generated set  grouped with respect to polyhedral unit combination. $A$-cations are omitted for clarity.
}
\label{generated_structures}
\end{figure}

\section{Results and Discussion}

Our exploration of OOV structures follows workflows used in other structure search algorithms, e.g., cluster expansions,\cite{ClusterExpansion,Zunger1990,Okhotnikov2016,VandeWalle/Ceder2002,VandeWalle2013,ICET,Lerch_uncle_2009}
because oxygen-deficient perovskites could be described as anionic alloys. 
Oxygen atoms and oxygen vacancies are mixed on the same site with the caveat that the OOV structures exhibit long-range order on the anion site. 
Thus, our investigation into the vacancy ordering principles is as follows:
($i$) We define the chemical space as $AB$O$_{2.5}$.
($ii$) We generate OOV structures with homogeneous vacancy patterns on (111)$_\mathrm{pc}$ layers based on observed configurations found in known OOV compounds \cite{Anderson1993}. 
Then, ($iii$) we identify ordering principles using these structures and DFT calculations.
Although we focus on \CAO, a nonmagnetic compound known\cite{CAO2008} to exhibit OOV phases [\autoref{fig:schematic}(b)], we also examine varying FBU-assemblies and employ other chemistries including SrFeO$_{2.5}$, SrMnO$_{2.5}$, LaNiO$_{2.5}$ to identify different contributions to the stability of the ordered vacancy pattern. 
Finally, ($iv$) we predict OOV structures with larger unit cells and demonstrate strategies for selecting compatible chemical identities to realize the designed OOV structure.

\subsection{Generated OOV structures}
\label{ss:generated_structures}

In \autoref{generated_structures}, we present examples sampled from the 50 OOV structures generated from our algorithm with varying combinations of FBU types.
In these structures, oxygen ligands are removed in various patterns, forming OOVs, while the corner connectivity of the FBUs are maintained as in pristine perovskites.
Notably, the structure set generated from our algorithm includes the prototypical examples of $AB$O$_{2.5}$ phases, demonstrating its effectiveness in capturing known and unknown patterns: 
No.\,1 corresponds to \autoref{fig:schematic}(f), No.\,2 to  \autoref{fig:schematic}(d), No.\,29 to  \autoref{fig:schematic}(b), and No.\,50 to  \autoref{fig:schematic}(c).
The complete list of space groups and FBU combinations of 50 generated structures are listed in
\supt{1}.

The advantage of applying a homogeneous planar vacancy pattern, rather than employing other structure enumeration schemes, is that this chemistry-informed constraint significantly reduces the number of possible arrangements.
The number of configurations is determined as a combinatorial, which rapidly increases with the size of unit cell \cite{Pickard2019,CALYPSO1}.
For example, in order to generate the 3D vacancy patterns with a unit cell size similar to those shown in \autoref{generated_structures}, requires assigning four vacancies in 24 oxygen sites, i.e., the combination $_{24}C_{4} =$ 10,626 structures for a single unit cell shape.
Owing to this rapid increase in the number of configurations from 3D sampling, many of which are unlikely to be low energy OOVs, techniques exploiting symmetry, genetic algorithms, or machine learning methods often need to be utilized to make the search tractable.\cite{Oguchi2018,Hautier/Ceder2011_DMSP,Sun/Ceder2017,Allahyari2020,Pretti_symmetry-based_2020,Collins/Rosseinsky2017,Lyakhov2013}
Our algorithm surveys vacancy arrangements within a 2D unit plane then employs translational symmetry constraints to efficiently generate the 50 OOV structures with varying polyhedral assemblies.

To the best of our knowledge, most previously identified  OOV phases have vacancy channels along either the [100]$_\mathrm{pc}$ or [110]$_\mathrm{pc}$ directions.
It is noteworthy that our structure generation algorithm completely encompasses structures with such vacancy channels, because our algorithm defines layer vacancy pattern on (111)$_\mathrm{pc}$ layer.
In a perovskite structure, the [100]$_\mathrm{pc}$ vector is neither orthorgonal nor parallel to all (111)$_\mathrm{pc}$ planes. On the other hand, a [110]$_\mathrm{pc}$ vector is parallel to half of the family of (111)$_\mathrm{pc}$ planes, and again neither orthorgonal nor parallel to the other half  of (111)$_\mathrm{pc}$ planes.
This relation means that when vacancy channels exist along either [100]$_\mathrm{pc}$ or [110]$_\mathrm{pc}$ direction, there are always unique (111)$_\mathrm{pc}$ planes that have an individual intersecting point for each vacancy channel.
In other words, a set of parallel vacancy channels (along [100]$_\mathrm{pc}$ or [110]$_\mathrm{pc}$ directions) maps to the same planar pattern on the (111)$_\mathrm{pc}$ planes by a specified translational vector.
Thus OOV structures following these principles can be completely enumerated by our algorithm.
As our algorithm surveys varying translational vectors, we are also able to obtain OOV structures both with and without vacancy channels.
Indeed, we find the formation of vacancy channels along low-index directions is closely related to the stability of the OOV phases (\emph{vide infra}).

\begin{figure}[t]
\centering
\includegraphics[width=0.85\columnwidth]{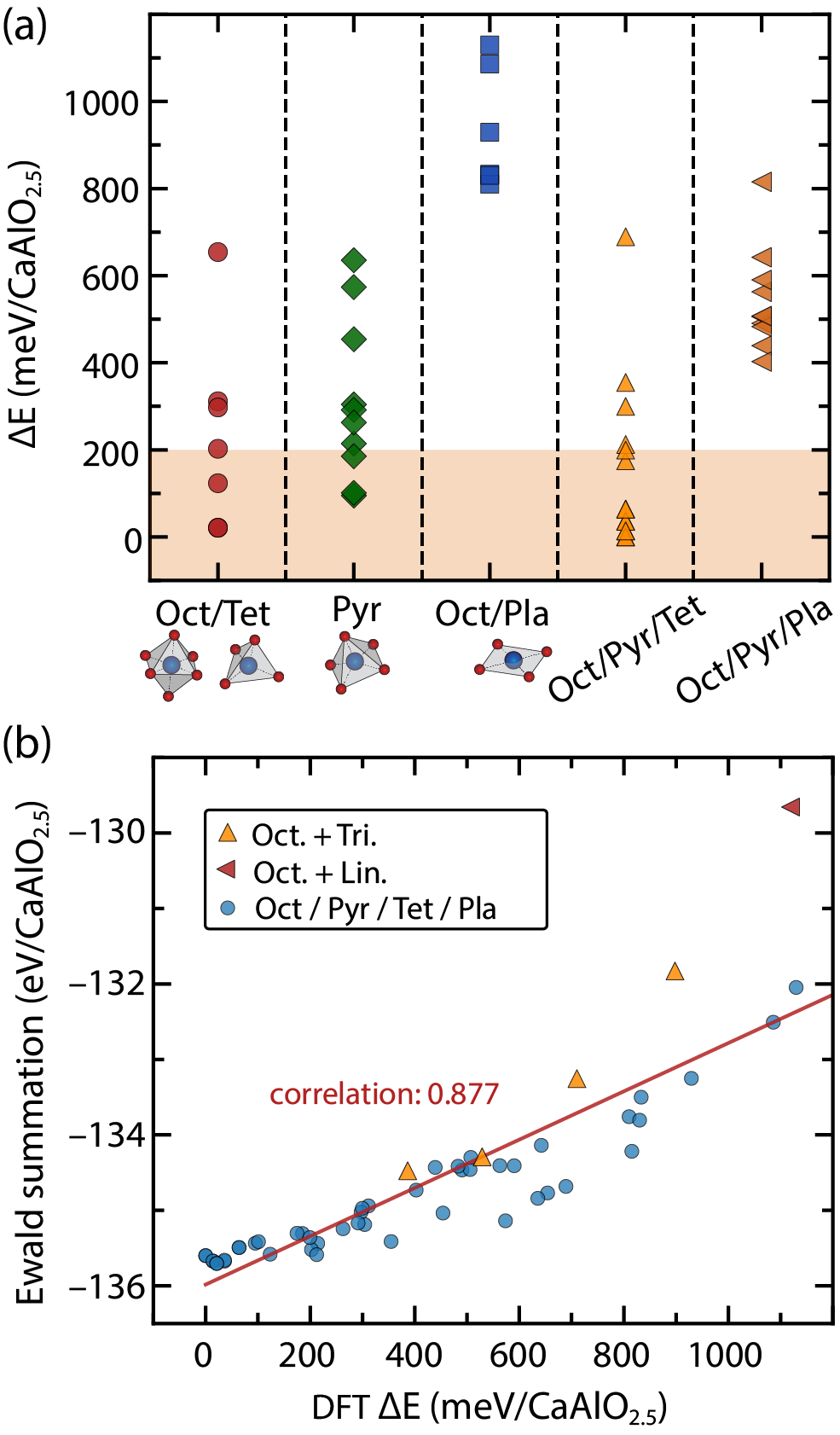}
\caption{(a) DFT calculated energies for CaAlO$_{2.5}$ OOV phases classified by constituent FBUs. Each structure is relaxed within the initial symmetry imposed by the OOVs.  The energetic range of synthesizability \cite{Sun/Ceder2016} is indicated as shaded area.
(b) The correlation between the DFT energies and Ewald summation (dipolar electrostatic energy using nominal ionic charges). Pearson correlation coefficient supports the linear correlation between the first-principles and surrogate energy models.
}
\label{fig:CAO_energies}
\end{figure}

\subsection{Ordering principles of OOV phases\label{ss:ordering_principles}}

\subsubsection{Preference for FBU-type \label{sss:preference_CAO}}

In \autoref{fig:CAO_energies}(a), we show the clustering of DFT calculated total energies for \CAO depends on the combination of polyhedra within generated OOV structure set.
\footnote{%
We note that the OOV structrures were relaxed within the initial symmetry imposed by hettotype OOV geometry without further symmetry breaking.
In some OOV structures such as  No.\,50 corresponds to the CaMnO$_{2.5}$-type structure,\cite{Caignaert1997} the initial symmetries are equivalent to the ground state space group.
On the other hand, No.\ 29 [\autoref{generated_structures}(b)] is the brownmillerite structure with $Imma$ space group,
which is known to exhibit various energy-lowering distortions.\cite{Young/Rondinelli2015,Parsons2009,Tian2018,Lu/Yildiz2020,Lu/Yildiz2016}%
When we applied the identified distortions to \CAO, we found the energy is further lowered by $\sim$236 meV/f.u.
}
In particular, structures comprising square planar units [Oct/Pla and Oct/Pyr/Pla in \autoref{fig:CAO_energies}(a)] have significantly higher energies compared to other groups.
The energetics are consistent with the chemical preference of 4-coordinate Al, which strongly prefers tetrahedral rather than square planar coordination \cite{Waroquiers/Hautier2017,Shannon1976}.
We confirm that such preference for a FBU-types depends on the $B$-cation by changing the cation identity.
Our analogous calculation for LaNiO$_{2.5}$ indicate that Ni is more stable in OOV structures comprising octahedral and square planar geometries [Oct/Pla] than other combinations of FBUs (see \supf{2}), 
which is consistent with the preference of Ni$^{2+}$  to adopt square planar coordination.\cite{Waroquiers/Hautier2017,Shin/Rondinelli2022a}

In \autoref{fig:CAO_energies}(b),
we show there is strong linear correlation between our DFT calculated energies and those obtained from an ionic model accounting for dipolar electrostatic interactions.
The energies in the ionic model were computed using an Ewald summation, by assuming nominal charge states (Ca$^{2+}$, Al$^{3+}$, and O$^{2-}$), resulting in a Pearson correlation coefficient of 0.877 with the DFT energies.
The correlation between the two methods becomes even higher (0.933) when we exclude triangular and linear coordinations [\autoref{generated_structures}(d-e)],
where the assumed charge states may not be maintained locally by the FBUs with the low coordination numbers.
This strong correlation suggests that the stability of OOV phases can be assessed via simpler ionic parameters, such as average bond lengths and bond valences, without needing to account for a more complete description of chemical bonding and electronic interactions by DFT.

We find that the energies within each group of FBU combinations can vary substantially [\autoref{fig:CAO_energies}(a)], which we attribute to elastic effects and explore further below.
We outline the argument briefly here. If the charge state of the $B$ cation is same, then FBUs with lower coordination numbers have shorter $B$--O bond lengths to compensate for the undercoordination compared to larger coordination polyhedra, e.g., octahedra.
Additionally, FBUs such as square pyramids and tetrahedra have nonorthogonal $B$--O bonds, which result in complex atomic relaxations in OOV phases\cite{Shin/Galli2023,Shin/Rondinelli2022a,Tian2018} to accommodate cooperative bond stresses.
Thus, understanding how atomic relaxations manifest and accommodate FBU-assemblies requires a description of both electrostatic and elastic interactions underlying the OOV. 

\begin{figure}[t]
\centering
\includegraphics[width=0.81\columnwidth]{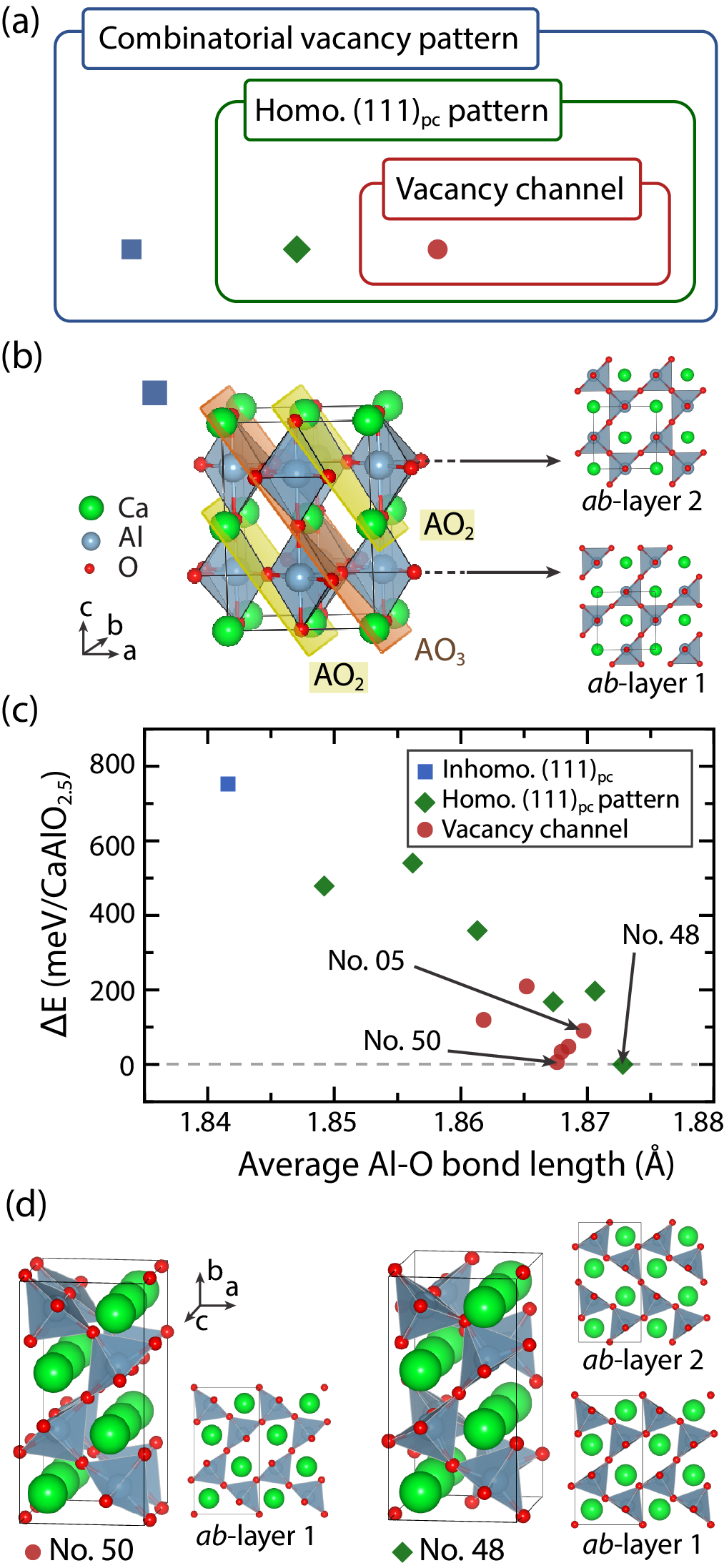}
\caption{(a) Venn diagram of the OOV structures generated with (i) combinatorial enumeration (blue square), (ii) with homogeneous (111)$_\mathrm{pc}$ layer patterns (green diamond), and (iii) with vacancy channels (red circles).
(b) Atomic structure of the pyramidal network generated with inhomogeneous densities on the  (111)$_\mathrm{pc}$ layers.
(c) Energies of CaAlO$_{2.5}$ square pyramidal networks as a function of the average Al--O bond length, relative to Structure No.\,48 whose $\Delta$E is 95 meV/\CAO in \autoref{fig:CAO_energies}.
(d) Two lowest-energy structures among the square pyramidal networks. No.\,48 and No.\,50 have identical vacancy patterns on the $ab$-plane, but differ by their translation of the in-plane pattern along the  stacking $\vec{c}$ direction.
}
\label{fig:diagram}
\end{figure} 

\subsubsection{Formation of vacancy channels}
\label{sssec:channels}

Two notable ordering tendencies found in OOV structures include the presence of (i) vacancy channels along either the [100]$_\mathrm{pc}$ or [110]$_\mathrm{pc}$ directions and (ii) homogeneous vacancy densities on (111)$_\mathrm{pc}$ planes.
As we explained earlier, 
the set of OOV structures generated by our algorithm, having the same (111)$_\mathrm{pc}$ layer patterns, is inclusive of those structures with vacancy channels.
These structures also belong to the structure set generated from a combinatorial enumeration.
The relation among these sets is depicted in \autoref{fig:diagram}(a).

In order to understand the relationship between phase stability and these classes,
we selected pyramidal networks with the \CAO chemistry for our primary analysis.
Within \ABO oxides, square pyramidal networks consist of a single FBU-type, the $B$O$_5$ square pyramid, 
whose oxide ligand positions remain close to $B$O$_6$ octahedron.
Thus, various assemblies may be realized through rigid unit distortions within the perovskite matrix without extreme deformation in polyhedral units, making the pyramidal networks a suitable group for investigating the effect of polyhedral assemblies on the stability of OOVs.

We examined 13 square pyramidal networks in total. Six of them have vacancy channels along [100]$_\mathrm{pc}$ while the other six structures have no vacancy channels but do exhibit homogeneous vacancy patterns on (111)$_\mathrm{pc}$ layers.
Additionally, we generated a \CAO structure with `nonuniform' vacancy densities on the (111)$_\mathrm{pc}$ layers to represent structures outside the scope of our structure generation algorithm [blue square in \autoref{fig:diagram}(a)]. Hereafter we refer to this structure as the nonuniform-(111)$_\mathrm{pc}$ structure.
As depicted in \autoref{fig:diagram}(b), the nonuniform-(111)$_\mathrm{pc}$ structure is distinct from other structures as it consists of alternatively stacked $A$O$_3$ and $A$O$_2$ layers along the [111]$_\mathrm{pc}$ direction.
It is noteworthy that the (100)$_\mathrm{pc}$ layer of this structure [Layer 1 and Layer 2 in \autoref{fig:diagram}(b)] share the same in-plane arrangement of pyramids as Structure No.\,5 [shown in \autoref{fig:sq_networks_energetics}(a)]. 
Given that Structure No.\,5 is experimentally observed\cite{Reller/Poeppelmeier1982}, we consider 
the vacancy ordering found in  the nonuniform-(111)$_\mathrm{pc}$ structure to be plausible.

In \autoref{fig:diagram}(c), we show the energy distribution of the pyramidal networks with respect to the  average Al--O bond length.
First, the energy distribution reveals that OOV structures with vacancy channels (red circles) tend to have low energies.
Among structures with a uniform (111)$_\mathrm{pc}$ pattern but without vacancy channels (green diamonds) some structures exhibit low energies, but the overall energy distribution is more dispersed, reaching up to $\Delta\mathrm{E}\sim$600\,meV/\CAO.
We note that the pyramidal network with the lowest energy belongs to this category [No.\,48  in \autoref{fig:diagram}(d)], while Structure No.\,50 [\autoref{fig:diagram}(d)] is only 6 meV/\CAO higher in energy. 
These two structures share the same $ab$-plane arrangement of pyramids, but in No.\,48, the $ab$-layers are shifted by $\frac{1}{2}\vec{b}$ upon stacking along $\vec{c}$.
This arrangement resembles the structure appearing at microdomain boundaries in SrMnO$_{2.5}$ \cite{Caignaert1986}.

In contrast, the nonuniform-(111)$_\mathrm{pc}$ structure [blue square in \autoref{fig:diagram}(c)] is extremely unstable with $\Delta$E of 753\,meV/\CAO.
This substantial energy difference is surprising, considering that this structure shares the same $ab$-plane arrangement with No.\,5 [\autoref{fig:sq_networks_energetics}(a)] whose $\Delta$E is only 90\,meV/\CAO.
The nonuniform-(111)$_\mathrm{pc}$ structure also has a shifted stacking of the $ab$-plane patterns, which does not necessarily lead to high energy when considering the similar energies of No.\,48 and No.\,50.
Although we only surveyed one structure with inhomogeneous vacancy densities on the (111)$_\mathrm{pc}$ layers, this extremely large $\Delta$E supports our hypothesis that inhomogeneous vacancy densities lead to unstable OOV structures.

\autoref{fig:diagram}(c) shows a clear correlation between total energy and the average Al--O bond length, with an average bond length close to 1.87\,\AA\xspace resulting in some of the lowest energy of \CAO pyramidal networks.
We also explored other descriptors associated with coordination environment, such as AlO$_5$ polyhedral volume, global instability index\cite{Salinas_GII1992}, and average bond valence of Al and Ca. These quantities result in a similar trend with  \autoref{fig:diagram}(c) as shown in \supf{5}.
This correlation indicates that the stability of OOV phases are determined by whether the pyramidal network allows for cooperative atomic displacements without frustrating the network 
to achieve optimal bond lengths.

\begin{figure}[t]
\centering
\includegraphics[width=0.85\columnwidth]{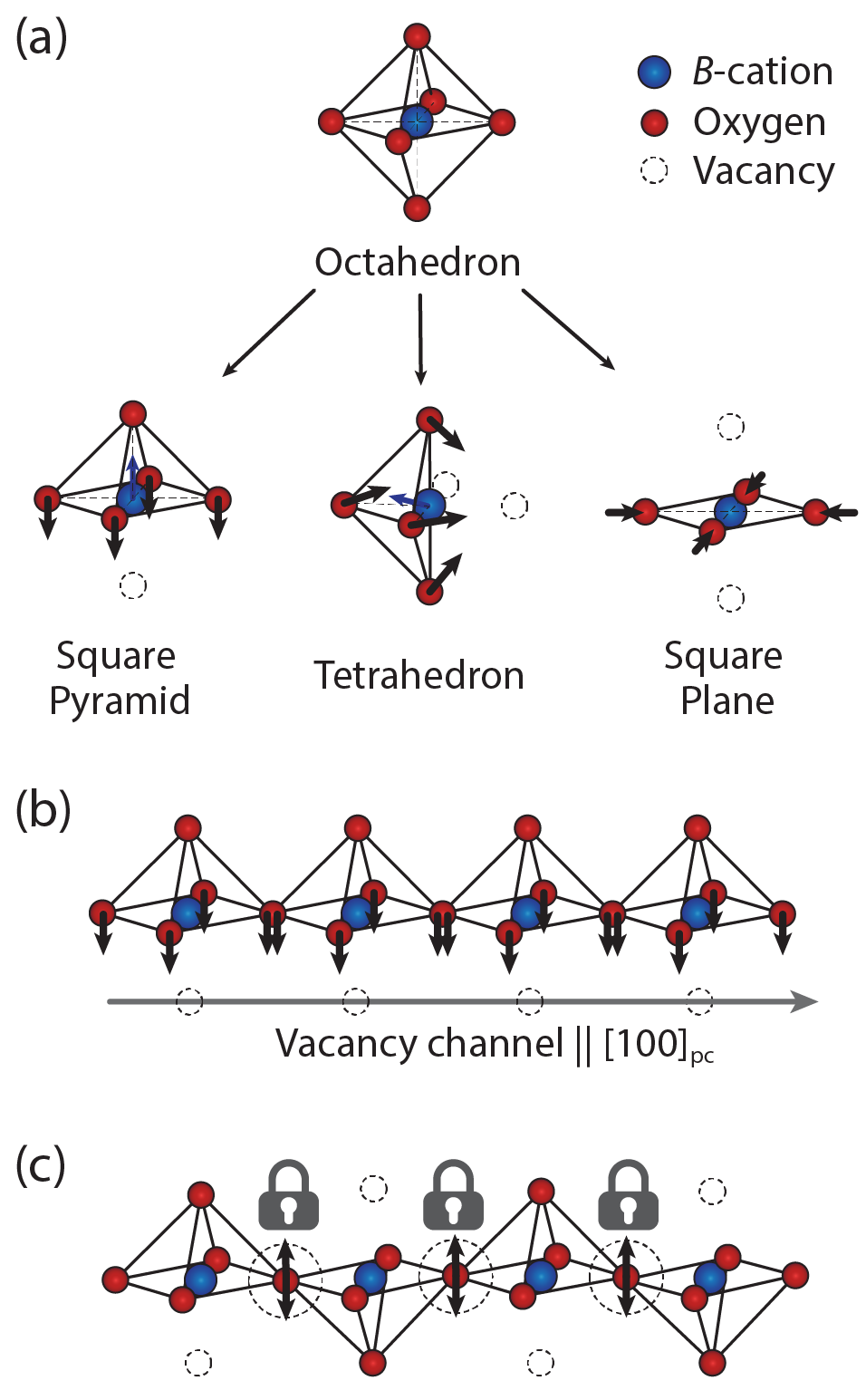}
\caption{(a) Schematic illustration of OOV-induced stresses acting on $B$-cation and oxygen ligands in square pyramidal, tetrahedral, and square planar geometries. (b) Stresses on bridging oxygen atoms in parallelly aligned $B$O$_5$ pyramids by forming vacancy channel along [100]$_\mathrm{pc}$ direction. Bridging basal oxygen atoms are displaced cooperatively toward the vacancy channel. (c) Anti-parallel alignment of $B$O$_5$ pyramids without forming a vacancy channel. Bridging oxygen atoms are locked by compensating stress components.
}
\label{fig:OOV_distortion}
\vspace{-1em}
\end{figure}

The perspective of atomic relaxation helps explain the energetic merit of forming vacancy channels.
Vacancy channels in pyramidal networks require that the AlO$_5$ pyramids are parallelly aligned because the apical oxgyen atom of each pyramid is positioned opposite to the vacancy site (\emph{trans} configuration).
This parallel alignment of pyramids is energetically favorable because it permits distinct apical and basal bond lengths within the pyramids.
For example in No. 50, AlO$_5$ has an apical bond length of 1.80 \AA, which is $\sim$0.1\AA\xspace shorter than its basal bonds (1.89 \AA).
When an AlO$_6$ octahedron transforms into an AlO$_5$ pyramid, the basal oxygen atoms displace toward the vacancy site and so that the Al than displaces as shown in \autoref{fig:OOV_distortion}(a) and \supf{3}.
For the parallel alignment of pyramids [as illustrated in \autoref{fig:OOV_distortion}(b)], two neighboring pyramids induce stresses on the bridging oxygen atom in the same direction,  cooperatively displacing the bridging anion.
On the other hand, when the pyramids are anti-parallelly aligned, 
the stress components on the bridging oxygen from two neighboring pyramids are compensated, locking the oxygen atom [\autoref{fig:OOV_distortion}(c)].
This locking results from tension between preferred Al--O bonds. 
Additionally, the exceptionally high energy of the nonuniform-(111)$_\mathrm{pc}$ structure in \autoref{fig:diagram}(b) can also be explained using these arguments.
Because all AlO$_5$ pyramids are anti-parallelly connected via basal oxygen atoms, the locked oxygen atoms create an elastic energy penalty and hence a higher energy phase.
Thus, the formation of vacancy channels are energy lowering because the cooperative distortions of $B$O$_5$ pyramids help to optimize the of bond valence of the $B$ cation in OOV structures.

The concept of cooperative distortions can also be applied to various FBUs like tetrahedra and square planes, because these FBUs tend to reorganize the oxygen ligands as shown in \autoref{fig:OOV_distortion}(a).
These FBUs with lower coordination number have shortened $B$--O bonds; in addition, the tetrahedral geometry further reforms the O--$B$--O bond angles to recover the ideal tetrahedral shape.
Because these FBUs are stiffer than octahedra, owing to the lower coordination number, we find octahedral units are more likely to host distortions in the OOV structures.
Understanding how the various FBUs accommodate the relaxations requires examining 
the internal strains across the bond networks (\emph{vida infra}).

Considering  that FBUs exhibit inherent displacements of oxygen ligands from octahedral geometries, we can explain the disadvantage of having inhomogeneous vacancy densities on (111)\PC layers. The (111)$_\mathrm{pc}$ layers in perovskites consist of close-packed $A$O$_3$ units.
Although FBUs in OOV structures tend to reorganize oxygen ligands and recover $B$O$_x$ shapes close to their ideal forms [\autoref{fig:OOV_distortion}(a)],
those oxygens are constrained in nominally close-packed (or eutactic) (111)$_\mathrm{pc}$ layers with either no or few vacancies.
This constraint makes it difficult to achieve the optimal bond valence of $B$-cations, increasing the energy of systems.

\begin{figure*}[t]
\centering
\includegraphics[width=0.82\paperwidth]{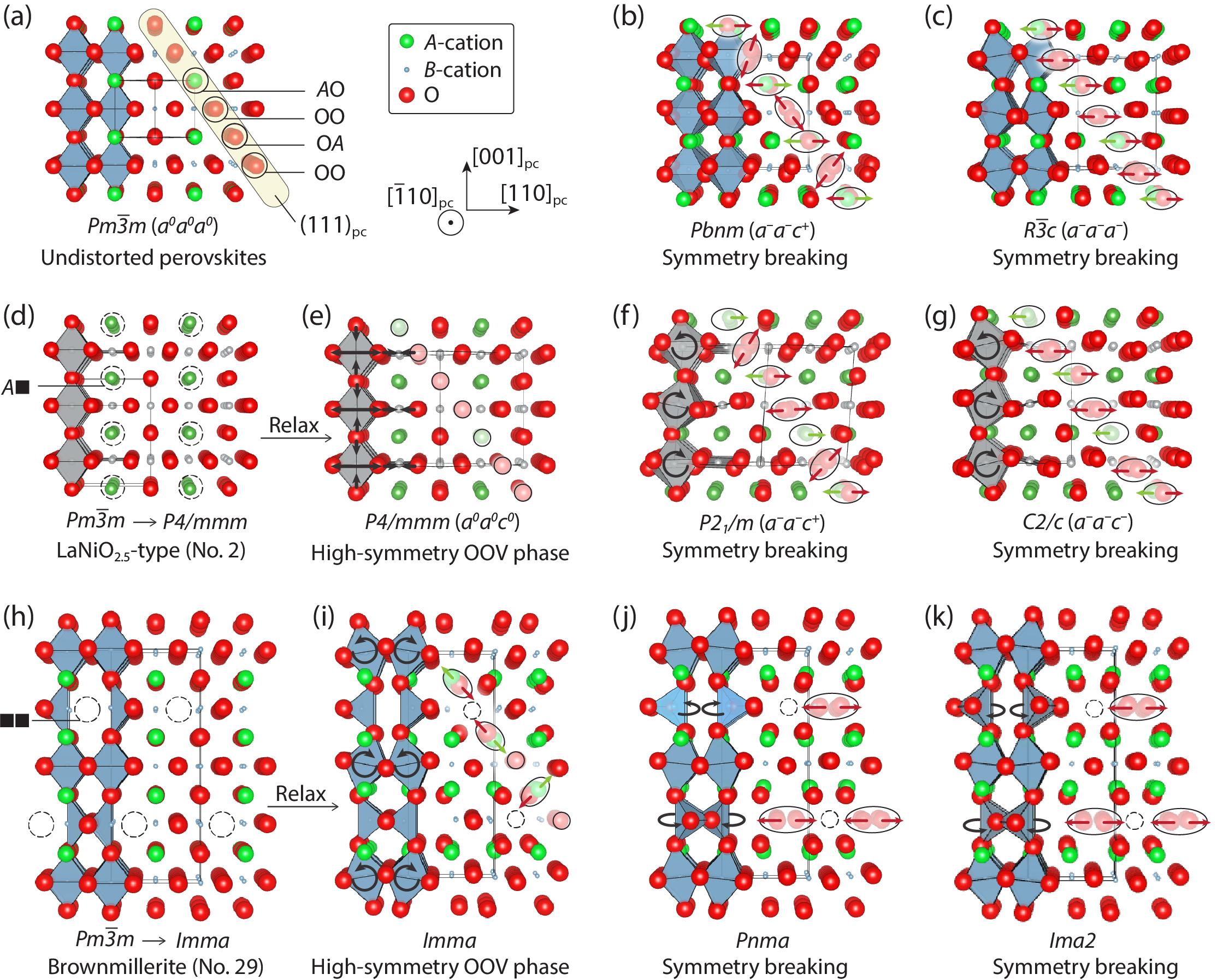}
\caption{Atomic structures of (a-c) pristine perovskites, (d-g) LaNiO$_{2.5}$-type (No.\ 2), and (h-k) brownmillerite (No.\ 29) structures as viewed from the $[\bar{1}10]_\mathrm{pc}$ direction. The leftmost column of OOV phases (d,h) shows vacancies formed in the cubic perovskite matrix,  while (e,i) show atomic relaxation within the symmetry imposed by OOVs. The right two columns illustrate typical cooperative distortions and symmetry breaking in each structure type. Atoms are scaled based on their ionic sizes, and displacements of atoms within one (111)$_\mathrm{pc}$ layer are highlighted with solid ellipses and arrows. Oxygen vacancy positions are annotated by dashed circles and labeled as black squares ($\blacksquare$). For LaNiO$_{2.5}$-type structures (d-g), green and gray spheres represent La and Ni, respectively.
}
\label{fig:110_view}  
\vspace{-0.5em}
\end{figure*}

\subsubsection{Role of $A$-cations in close-packed $A$O$_{3-\delta}$ lattice}
\label{ss:A_cation_role}

While we have primarily focused on the bond valence of $B$-cations to explain the stability of OOV phases, it is essential to note that the $A$-cation also play a significant role in determining the stability of OOV phases. This is because $A$-cations also comprise the close-packed planes of the perovskite structure and their bond valences need to be optimized to ensure stability. 
To elucidate the role of the $A$-cations in the atomic relaxations found in OOV structures, we compare the typical distortions appearing in pristine perovskites and two OOV phases (LaNiO$_{2.5}$-type and brownmillerite structures) in \autoref{fig:110_view}.

In \autoref{fig:110_view}(a-c), we show the prototypical distortions appearing in cubic perovskites that arise from small $A$ cations.
When the Goldschmidt tolerance factor\cite{Goldschmidt1926} is smaller than 1, the cubic perovskite structure [\autoref{fig:110_view}(a)] with $Pm\bar{3}m$ space group and $a^0a^0a^0$ Glazer pattern \cite{Glazer1972} distorts 
into orthorhombic (\autoref{fig:110_view}(b); $Pbnm$; $a^-a^-c^+$) and rhombohedral (\autoref{fig:110_view}(c); $R\bar{3}c$; $a^-a^-a^-$) phases.
These distortions involve octahedral rotation and tilting motions, which compensate bond valence of undercoordinated $A$-cations \cite{Benedek2013}.
When displaying atomic structures of perovskites along [110]\PC direction, as shown in \autoref{fig:110_view}(a-c), these distortions can be characterized as transverse acoustic motions of $A$O- and OO-chains where neighboring atoms in a chain displace to the opposite transverse directions.
These distortions can be interpreted with respect to the 
cubic lattice with nominally 
$A$O$_3$ close-packed planes and $B$-cations occupy octahedral interstitial sites. 
When a small $A$-cation is undercoordinated, 
motions of $A$O- and OO-chains tune the $A$--O bond lengths, optimizing the bond valence of $A$-cations.

OOV structures like No.\,2 (LaNiO$_{2.5}$-type) and No.\,29 (brownmillerite) exhibit similar transverse distortions of the $A$O- and OO-chains when viewed along certain [110]\PC directions. 
We find that in some [110]\PC directions, longer wavelength distortions occur. 
However, to have a better comparison with pristine perovskites, we examine the [110]\PC direction with the shortest wavelength distortions. 
We also provide the [110]\PC view of pyramidal networks in \supf{4}, 
which also shows the acoustic motions of $A$O- and OO-chains.
First, the LaNiO$_{2.5}$-type structure does not show a distortion of $A$O- and OO- chains when relaxed without symmetry breaking [$P4/mmm$; \autoref{fig:110_view}(d,e)].
The notable change in structure is a compressive Jahn-Teller distortion to the octahedra, which occurs due to internal strain induced by square planar units.\cite{Shin/Rondinelli2022a}
When symmetry-breaking distortions are applied, there are two competing phases close to the ground state, $C2/c$ and $P2_1/m$,\cite{Alonso1995,Moriga1994,Shin/Rondinelli2022a} which resemble the rotation patterns found in perovskites. 
However, while the $C2/c$ phase [\autoref{fig:110_view}(g)] shares nearly identical displacement patterns found in $R\bar{3}c$ perovskites [\autoref{fig:110_view}(c)], we find that the $P2_1/m$ phase [\autoref{fig:110_view}(f)] realizes different transverse orientations of OO-chains compared to those in $Pbnm$ perovskites [\autoref{fig:110_view}(b)].

In the brownmillerite structure, distortion modes are observed during the atomic relaxation within the high symmetry phase ($Imma$), as depicted in \autoref{fig:110_view}(h) and (i).
While this relaxation is often interpreted as the elongation and rotation of octahedra\cite{Zhang/Galli2020},
the same structure relaxation can be interpreted using the motions of $A$O-chains where, where oxygen atoms displace toward the vacancy channel ($\blacksquare\blacksquare$) and $A$-cations displace in the opposite direction.
(We do not find electrostatic interaction to be the primary driving force behind these distortions, at least in general OOV structures. In brownmillerites, $A$O-chains displace with oxygen atoms. They move toward the vacancy position and the $A$-cation moves away from the vacancy position, but this trend is not observed in the LaNiO$_{2.5}$-type structure.)
This motion displaces the $A$-cations closer to other oxygen atoms, supplementing the bond valence undercoordinated $A$-cations when oxygen vacancies are present.

The symmetry breaking to achieve the brownmillerite structure involves the twisting of tetrahedral chains \cite{Young/Rondinelli2015,Tian2018,Parsons2009}.
The associated motion occurs in the OO-chain [\autoref{fig:110_view}(j,k)], corresponding to the equatorial oxygen atoms forming the tetrahedra.
Interestingly, this motion in the OO-chain alters the $A$O-chains in that the transverse direction points toward the vertical direction and the amplitude of the motion decreases.
This rearrangement  of chain displacements can be understood as a route to reduce the elastic penalty. 
The displaced oxygen atoms in the OO-chains supplement the bond valence of $A$-cations, and the motion of the $A$O-chains is reduced which lower the energy penalty from their displacements. 
Although the two symmetry-breaking distortions in \autoref{fig:110_view}(j,k) differ in that they stabilize antipolar and polar phases with twisting of tetrahedral chains, the structures share qualitatively the same chain motions.

The similarity of motions along the [110]\PC direction demonstrates that
atomic relaxations in pristine perovskites and OOV structures are both realized within 
the $A$O$_3$ close-packed planes.
The distinction between pristine perovskites and OOV structures lies in the fact that  
$A$O- and OO-chains adopt different patterns 
in the presence of oxygen vacancies.
This tendency reinforces the advantages of having homogeneous vacancy densities on the (111)\PC layers.
If vacancy densities are inhomogeneous among the (111)\PC layers, then one (111)\PC layer without vacancies may exhibit a certain motion that may be incompatible with another (111)\PC layer with oxygen vacancies, leading to an unstable and unlikely atomic structure.

Additionally, the merit of forming a vacancy channel along low-index crystallographic directions can be understood from the fact that $A$-cations repeat in space with $\langle100\rangle$\PC vectors in perovskites.
When oxygen vacancies are present around an $A$-cation, the surrounding atoms are displaced to minimize the formation energy of the local vacancy.
Because these atomic distortions repeat on the \textit{next} $A$-cation due to acoustic motion in close-packed lattice,
the same relative position to the next $A$-cation becomes the preferable vacancy site,
making oxygen vacancies repeat along the acoustic chain direction. 
The same interpretation applies to $B$-cations, because the $B$-cations also repeat with $\langle100\rangle$\PC vectors.
Consequently, the formation of vacancy channels along low-index crystallographic directions are preferable to mutually achieve optimal bond valences of the $A$- and $B$-cations in OOV structures.

\subsubsection{Use Cases: FBU-assemblies with vacancy channels}
\label{ss:case_studies}

Although the formation of vacancy channels is a desired feature to stablize OOV structures,
allowing better atomic relaxation of $A$- and $B$-cations,
there are various possible polyhedral assemblies compatible with vacancy channels.
In order to investigate which assemblies lead to low-energy OOV structures,
we examine elastic penalities in pyramidal networks (Pyr group), structures with both octahedra and tetrahedra (Oct/Tet group), and both octahedra and square planes (Oct/Pla group).

\begin{figure}[t]
\centering
\includegraphics[width=0.81\columnwidth]{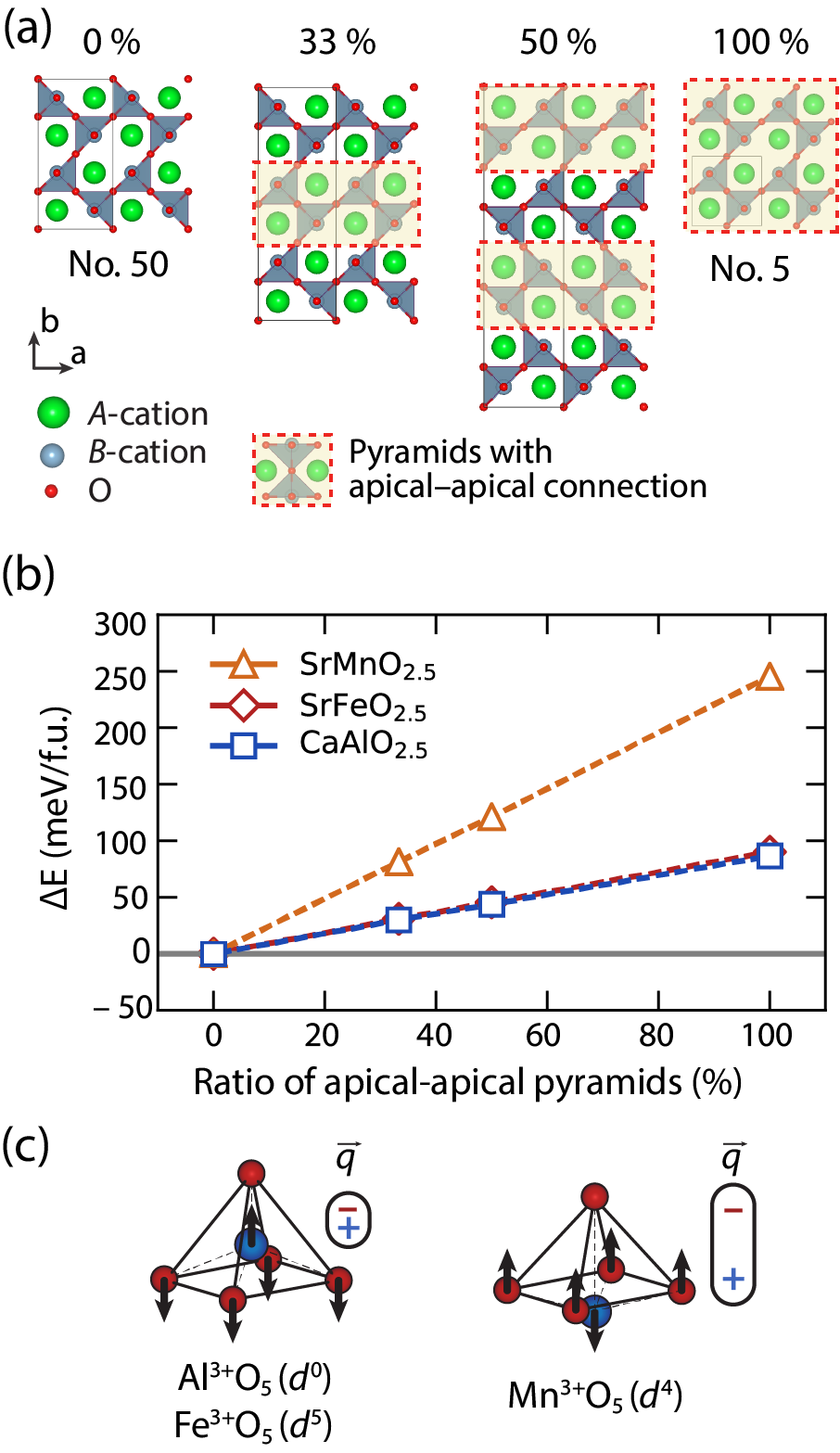}
\caption{(a) Square pyramidal networks with [100]$_\mathrm{pc}$ vacancy channels. These structures have different ratios of pyramidal units with apical--apical connections (0, 33, 50, and 100 \%). In each structure, the pyramids with apical-apical connections are highlighted, and corresponding ratios are labeled.
(b) Energy differences in CaAlO$_{2.5}$, SrFeO$_{2.5}$, and SrMnO$_{2.5}$ as a function of the ratio in (a), referenced to the energy of No.\ 50 [$\Delta$E of \CAO is 101 meV/f.u.\ in \autoref{fig:CAO_energies}(a)]. Magnetic orders were assigned following the Goodenough-Kanamori-Anderson (GKA) rules \cite{Goodenough1963,Kanamori1959}, which are followed in the CaMnO$_{2.5}$-type structure.\cite{Shin/Rondinelli2021,Shin/Rondinelli2020a}
(c) Schematic illustration of dipole moments in a $B$O$_5$ square pyramid unit. Al and Fe atoms displace into the pyramid and reduce the dipoles, whereas Mn displaces below the basal plane and increases the dipole moment.
}
\label{fig:sq_networks_energetics}
\vspace{-1em}
\end{figure}

\paragraph{Square pyramid group.}
We show four square pyramidal networks in \autoref{fig:sq_networks_energetics}(a) that have vacancy channels along  the [100]$_\mathrm{pc}$ direction.%
\footnote{%
We note that the two structures labeled with 33 and 50 \% in \autoref{fig:sq_networks_energetics}(a) are not covered in our set of 50 structures because their large unit cell sizes cannot be captured in our surveyed (111)$_\mathrm{pc}$ plane.
These structures would be included if a longer $A$O$_3$ (111)$_\mathrm{pc}$ plane was used as input.
}
These experimentally observed structures\cite{Reller/Poeppelmeier1982} are differentiated by the in-plane connections. 
The connections can be classified by apical--apical, apical--basal, and basal--basal connections.
We find that the ratio of pyramids having apical-apical connections serves as a descriptor to distinguish the four structures. 
The highlighted area in \autoref{fig:sq_networks_energetics}(a) indicates which pyramids in each structure exhibits such connectivity amounting to 0, 33.3, 50, and 100 \% with respect to this descriptor.

Structure No.\ 50 with 0 \% apical--apical connection is the equilibrium structure of SrMnO$_{2.5}$\cite{Caignaert1986}. SrFeO$_{2.5}$ adopts the same structure under high pressure\cite{Zhu2014} despite exhibiting the brownmillerite structure (No.\ 29) at ambient pressure \cite{Young/Rondinelli2015}.
We find that the apical-apical connections descriptor exhibits an almost linear dependence with the energies of the structures, as shown in $\Delta$E in \autoref{fig:sq_networks_energetics}(b).
This energetics can be explained from both elastic and electrostatic aspects, because the $B$O$_5$ pyramid unit is polar by construction 
[$C_{4v}$ symmetry; \autoref{fig:sq_networks_energetics}(c)].
From an elastic point of view, the distinct apical/basal Al--O bond lengths in the pyramids play a crucial role. 
To be compatible with the pseudocubic lattice of a perovskite network, the pyramidal connections with pairing of long-basal and short-apical bonds reduces elastic instability.
On the other hand, from an electrostatic point of view, an apical--apical connection places two negative dipole charges at the bridging oxygen positions.
Thus, the in-plane arrangement of No.\,50 is advantageous because all apical oxygen atoms of all pyramids are connected to basal oxygen atoms of the other pyramids.
The competing stability of No.\,48 in \autoref{fig:diagram}(c) can be also explained by considering the same in-plane pyramidal arrangements.
Despite the different structure, this zigzag arrangement is found in infinite-layer CaCoO$_2$ \cite{Kim/Hwang2023}; here Jahn-Teller distorted square planes are also connected in a zigzag manner, which resembles the ice rule minimizing electrostatic interactions \cite{Pauling_IceRule1935,Williams1993molecular}.

We note that the elastic and electrostatic instabilities are correlated and balanced by tuning the position $B$-cation in the pyramid.
If the electric dipole is too large, the $B$ cation displaces closer to the center of pyramid to reduce the dipole.
On the other hand, if placing the $B$ cation at the center of the pyramid requires significantly different apical/basal bond lengths, then a smaller dipole moment will form to reduce the elastic cost from the elongated and compressed bonds.
Indeed, a AlO$_5$ pyramid in structure No.\,50 exhibits apical bond length of 1.80\,\AA\xspace and an average bond length of 1.89\,\AA. The position of the Al is 0.07\,\AA\xspace away from the center of pyramid and toward the basal plane, forming a dipole moment of 0.81\,Debye when calculated using nominal charges.

We further surveyed the energetics of the \SFO and \SMO pyramidal networks in \autoref{fig:sq_networks_energetics}(b), because the apical/basal bond ratio is dependent on the $d$-orbital filling\cite{Shin/Rondinelli2021,Shin/Rondinelli2020a} and different electric dipole moments are expected to result.
Interestingly by using nominal charges, we find that the dipole moment for FeO$_5$ is 1.27\,Debye, and the maximum $\Delta$E in \SFO with 100\% apical--apical connections is 90.1 meV/\SFO, similar to the increase in \CAO (84.2 meV/\CAO) between 0 and 100\% of apical--apical connections.
On the other hand, the dipole moment of MnO$_5$ is 3.46\,Debye and the maximum $\Delta$E increases by 245\,meV/f.u.\ in SrMnO$_{2.5}$.
This additional increase in $\Delta$E in \SMO can be attributed to the larger electrostatic penalty, because Mn$^{3+}$ has a $d^4$ electronic configuration and this results in a longer apical bond compared to the basal bonds \cite{Shin/Rondinelli2020a,Shin/Rondinelli2021}. Because this bond length difference displaces Mn below the basal plane, a larger electrical dipole moment is formed within the  MnO$_5$ pyramid.

\begin{figure}[th]
\centering
\includegraphics[width=0.85\columnwidth]{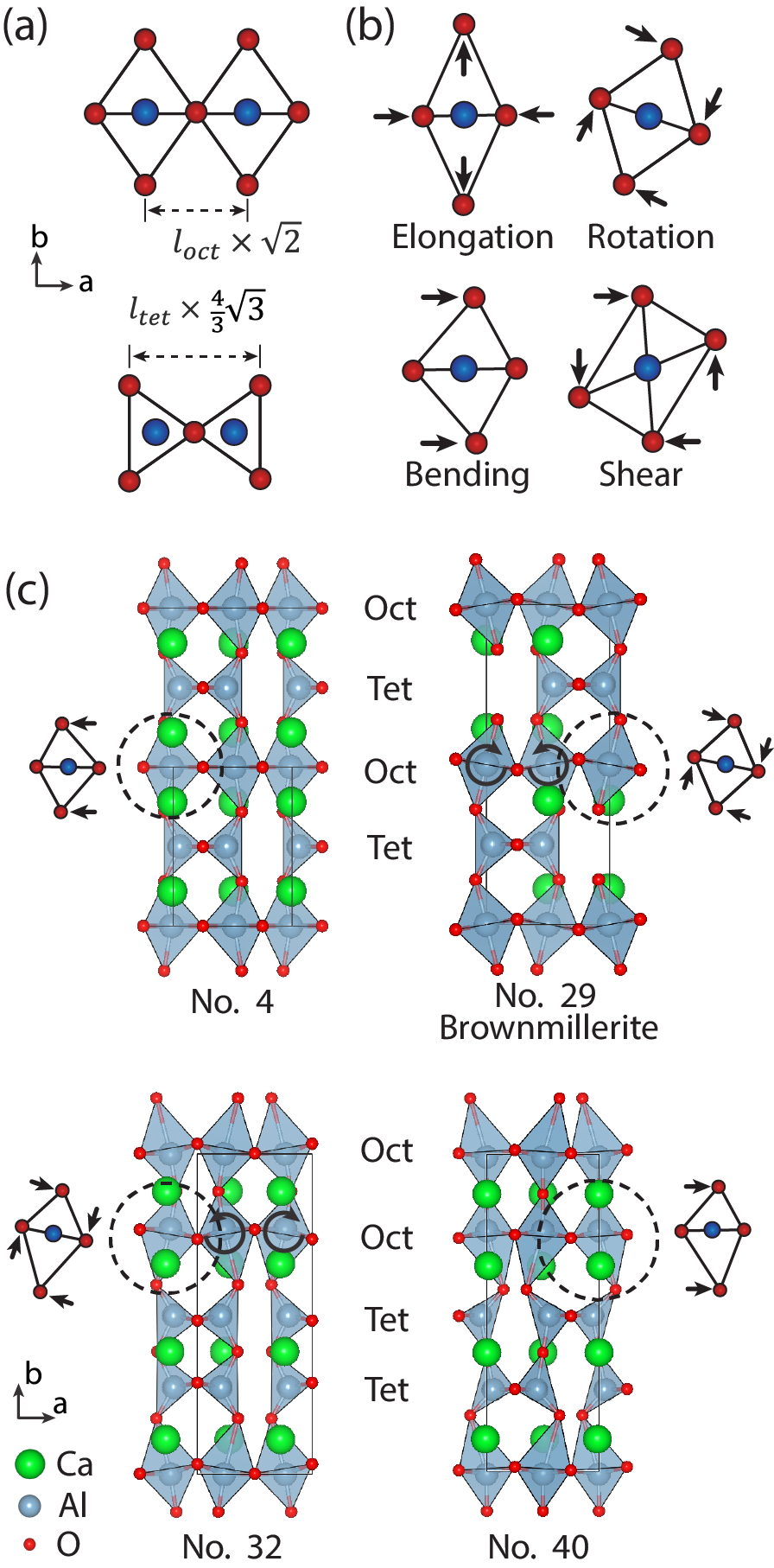}
\caption{(a) Projected distance between bridging oxygen atoms in octahedral and tetrahedral layers with ideal polyhedral geometries. Note that the atoms and distances are projected on to the $ab$-plane and $a$-direction, respectively.
(b) Prototypical octahedral distortions found in OOV structures. (c) Atomic structures of OOV patterns with octahedral and tetrahedral layers: No.\ 4, No.\ 29, No.\ 32, and No.\ 40. Distortions induced in the octahedral units found in each structure are indicated.
}
\label{fig:BM_distort}
\vspace{-1em}
\end{figure}

\paragraph{Structures with octahedra and tetrahedra.}
In our dataset, there are four OOV structures consisting of corner-connected octahedral and tetrahedral layers, No.\,4, 29, 32, and 40, as shown in \autoref{fig:BM_distort}(c).
The structures are distinguished by their stacking sequence of homogeneous octahedral and tetrahedral layers along $\vec{b}$ and the shift of the tetrahedral chains along $\vec{a}$, as listed in \autoref{table:BM}.
The most stable OOV pattern (No.\,29) exhibits displaced tetrahedral chains along the stacking direction and the octahedral/tetrahedral layers are alternatively stacked.
Structure No.\,4 differs from No.\,29 only by the tetrahedral chain shift such that the chains are vertically aligned.
Likewise, structures No.\,32 and 40 differ in between their polyhedral layers in every two perovskite layers; they are also distinguished by shifted and aligned tetrahedral chains.

The atomic relaxations found in structures within the Oct/Tet group can be understood by considering that the octahedral and tetrahedral layers have different O--O spacing between their apical oxygen atoms, as illustrated in \autoref{fig:BM_distort}(a).
When ideal polyhedral units are assumed, the O--O spacing along the $a$ direction is $l_{oct}\times\sqrt{2}$ for octahedral layers and $l_{tet}\times\frac{4}{3}\sqrt{3}$ for tetrahedral layers.
For example in structure No.\,29, the average Al--O bond lengths of the octahedra and tetrahedra are 1.97 and 1.74\,\AA, respectively, and the corresponding O--O spacing distance is 2.79 and 4.02\,\AA, respectively.
To achieve the corner-connectivity between these two mismatched layers, the apical oxygen atoms in the octahedral layers need to be further spaced (along the $a$-direction).
This is accomplished by distortions to the octahedral units rather than the tetrahedral  units, because the bonds in the tetrahedra are stiffer owing to lower coordination number.
Thus , octahedra are more susceptible to deformations, e.g., Jahn-Teller-like elongation/compression, rotation, bending, and shearing modes, as depicted in \autoref{fig:BM_distort}(b).
Deformation of the octahedra is a useful proxy of the stability in perovskite-related structures\cite{Guzman-Verri2019}, where the rotation/tilting and volume-conserving elongation/compression modes are energetically favorable distortions. In contrast, bending and shearing octahedral modes considerably increase the energy.

\begin{table}[t]
\caption{\label{table:BM}
Polyhedral connectivities and relative energies of OOV structures appearing in \autoref{fig:BM_distort}(c). The networks differ by the polyhedral stacking sequences (along $\vec{b}$) and the alignment of the tetrahedral chains (along $\vec{a}$).
The unit of energy difference ($\Delta$E) is meV/CaAlO$_{2.5}$, referenced to No.\,29 whose $\Delta$E is 21 meV/\CAO in \autoref{fig:CAO_energies}(a).
}
\centering
\begin{tabular}{l l l r }
\toprule
Structure    & Stacking sequence   & Tet. chains  & $\Delta$E \Bstrut \\ 
\midrule
No.\,4       &  Oct/Tet/Oct/Tet      & Aligned  &   181        \\  
No.\,29      &  Oct/Tet/Oct/Tet     &  Shifted  &    0        \\  
No.\,32      &  Oct/Oct/Tet/Tet     & Aligned   &    276       \\  
No.\,40      &  Oct/Oct/Tet/Tet     & Shifted   &    633      \\   
\bottomrule
\end{tabular}
\end{table}

While we find that octahedral elongation is found in all structures in \autoref{fig:BM_distort}(c), we observe that the stacking sequence determines whether the octahedral elongation is symmetric or asymmetric. The shift of the tetrahedral chains (along $\vec{a}$) determines whether the octahedra undergo rotation or bending distortions.
In structure No.\,29, the mismatch in the O--O spacing is accommodated by octahedral rotations whereas the octahedra in structure No.\,4 undergo bending. This difference leads to an 180\,meV/f.u.\ energy difference, as shown in \autoref{table:BM}.
Because structure No.\,32 and No.\,40 have an additional octahedral layer between the tetrahedral layers,  we observe considerable asymmetric octahedral elongation and reversal in the distortions.%
\footnote{%
We find this observation is also observed in the $AB$O$_{2.67}$ Grenier structures where two tetrahedral layers are separated by two octahedral layers\cite{Shin/Galli2023,Battle1990,Grenier1,Grenier2,Grenier3,Inkinen/Dijken2020}.
In this structure, the tetrahedral chains are vertically aligned as that arrangement is compatible with rotations of two octahedral layers, which is similar to the octahedral rotations found in No.\,32 but with a single tetrahedral layer.
}
Structure No.\,32 with vertically aligned tetrahedral chains exhibits octahedral rotations and structure No.\,40 with shifted chains develops bending distortions.
Again, the structure with the rotational mode is more energetically favorable, but both structure No.\,32 and No.\,40 have high $\Delta$E values of 276 and 633 meV/\CAO, possibly due to the asymmetric elongation.
Structure No.\,40 also exhibits deformed tetrahedral units, owing to unmatched O--O distances between the tetrahedral layers, making structure No.\,40 further unstable.

\begin{figure}[t]
\centering
\includegraphics[width=0.90\columnwidth]{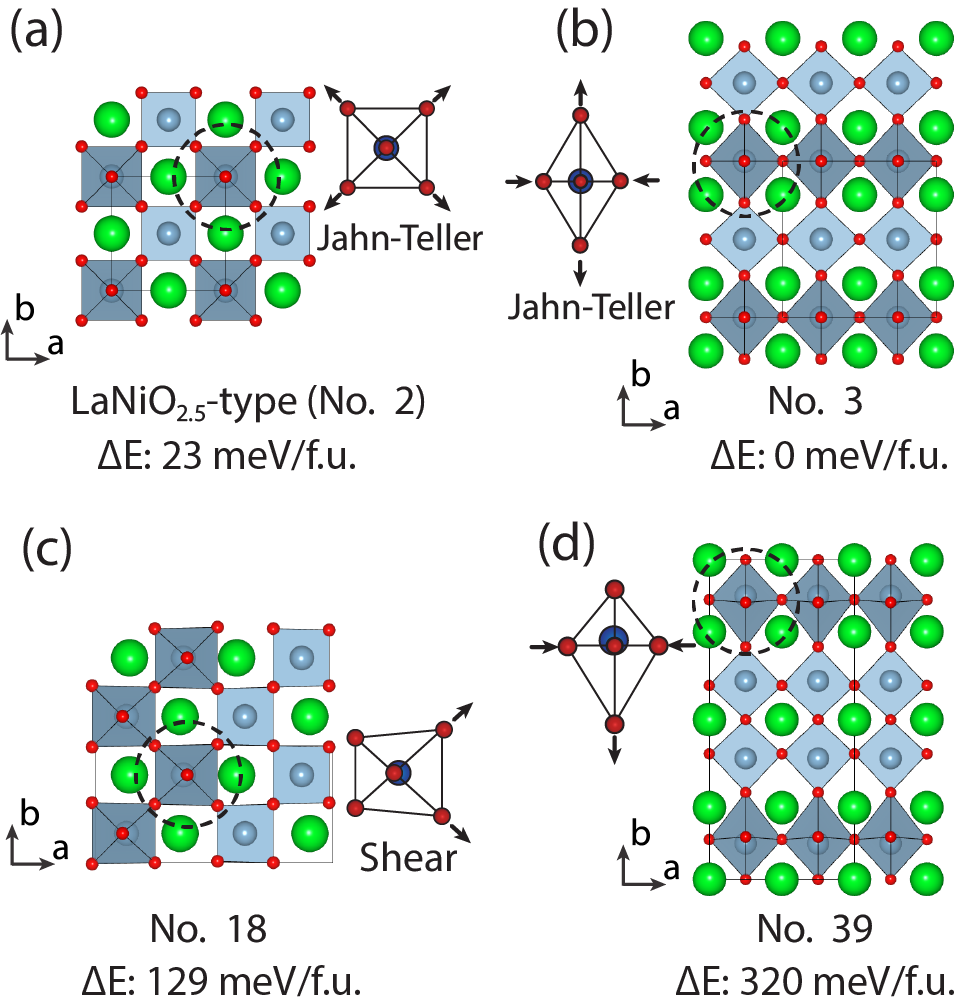}
\caption{OOV structures with octahedra and square planes; (a) No.\ 2, (b) No.\ 3, (c) No.\ 18, and (d) No. \ 39. Relative energies are referenced to structure No.\ 3 whose $\Delta$E is 810 meV/\CAO in \autoref{fig:CAO_energies}, and induced distortions to the octahedra on the $ab$-plane are depicted for clarity. Note that the Jahn-Teller distortions in the octahedra are compressive, axially along the $c$ direction, in No.\ 2 and tensile, axially along the $b$ direction, in No.\ 3.
}
\label{fig:Sq_Planar}
\end{figure}

As discussed earlier (see \autoref{fig:110_view}), 
the relaxation in brownmillerite involves oxygen atoms displacing toward vacancy sites and $A$-cation displacements away from the vacancy site.
While other structures in \autoref{fig:BM_distort}(c) also exhibit similar motions of among the $A$O-chains, we find that the center position of the $A$O layers remain equally spaced along the stacking direction (see \supf{6}). 
This finding further supports the view that relaxations in the OOV structures occur within closed pack framework of perovskite.

\paragraph{Structures with octahedra and square planes.}
There are four structures, No.\,2, 3, 18, and 39, in our dataset belonging to the Oct/Pla group.
These structures exhibit variations in the arrangement of octahedra and square planar units in the $ab$-plane.
Although Al$^{3+}$ is not energetically compatible with a square planar coordination, we use it in these coordination environments to make relative energy comparisons in \autoref{fig:Sq_Planar} referenced to Structure No.\,3.
\footnote{
Owing to the inherent instability of square planar coordiction for Al, we find the square planar units in Oct/Pla group structures converge to tetrahedral coordination when we artificially break the symmetry with random perturbations. This relaxation significantly lowers the energies of each structure by 336--738 meV/\CAO (see \supf{7}).
}
We find the distortions in these structures are primarily induced by the internal strain between the octahedra and square planar units.
When comparing the average Al--O bond length in structure No.\,2 (LaNiO$_{2.5}$-type structure), we find the average Al--O bond lengths of octahedra and square planes are 1.92 and 1.81 \AA, respectively.
Because an Al--O bond within a square plane is shorter and stiffer than that in an octahedra, the bridging oxygen atom between the two FBU types are stretched toward the square planes.
As a result, octahedra are subjected to different distortions as shown in \autoref{fig:Sq_Planar}; structures No.\,2 and 3 exhibit compressed and elongated Jahn-Teller-like  distortions of octahedra. 
Octahedra in structure No.\,18 are sheared asymmetrically, because  oxygen atoms connected to square planes are particularly stretched. 
Similarly, structure No.\,39 shows asymmetric elongation of octahedra.

Although structure No.\,2 is the experimentally known OOV structure for LaNiO$_{2.5}$,\cite{Shin/Rondinelli2022a} we find structure No.\,3 is more stable than No.\,2 by 23 meV/\CAO.
Typically, connecting two short bonds would generally lead to higher energy OOV structures.
However, in the case of structure No.\,3, the internal strain is resolved through a Jahn-Teller distortion of the octahedra.
This Jahn-Teller distortion elongates octahedra along the $b$ direction in \autoref{fig:Sq_Planar}(b), whereas the shorter Al--O bonds occurring along the $a$- and $c$-directions become compatible with the shorter average bond length of the square planar units.
This linear arrangement of square planes is often found in cuprates \cite{Huang/Lee2018,Park/Snyder1995}.
Thus, we anticipate the exploration of symmetry breaking in structure No.\,3 would be intriguing for LaNiO$_{2.5}$ or similar chemistries because the ground state structure No.\,3 may energetically compete with No.\,2.

On the other hand, structure No.\,18 and 39 exhibit asymmetric shearing and elongation, respectively, which results in higher energies.
When compared to the energy differences in structures of the Oct/Tet group, ranging from 181 to 633 meV/\CAO, we find these structures with asymmetric deformation are less unstable.
In contrast to the AlO$_4$ tetrahedral unit, which has an average Al--O bond length of 1.74\,\AA\xspace and nonorthogonal interbond angles, square planar units have a longer average Al--O bond length of 1.81\,\AA\xspace and nearly orthogonal bond angles.
We expect these differences decrease the elastic penalties when connected to the octahedra with longer Al--O bond lengths.

\begin{figure}[t]
\centering
\includegraphics[width=0.90\columnwidth]{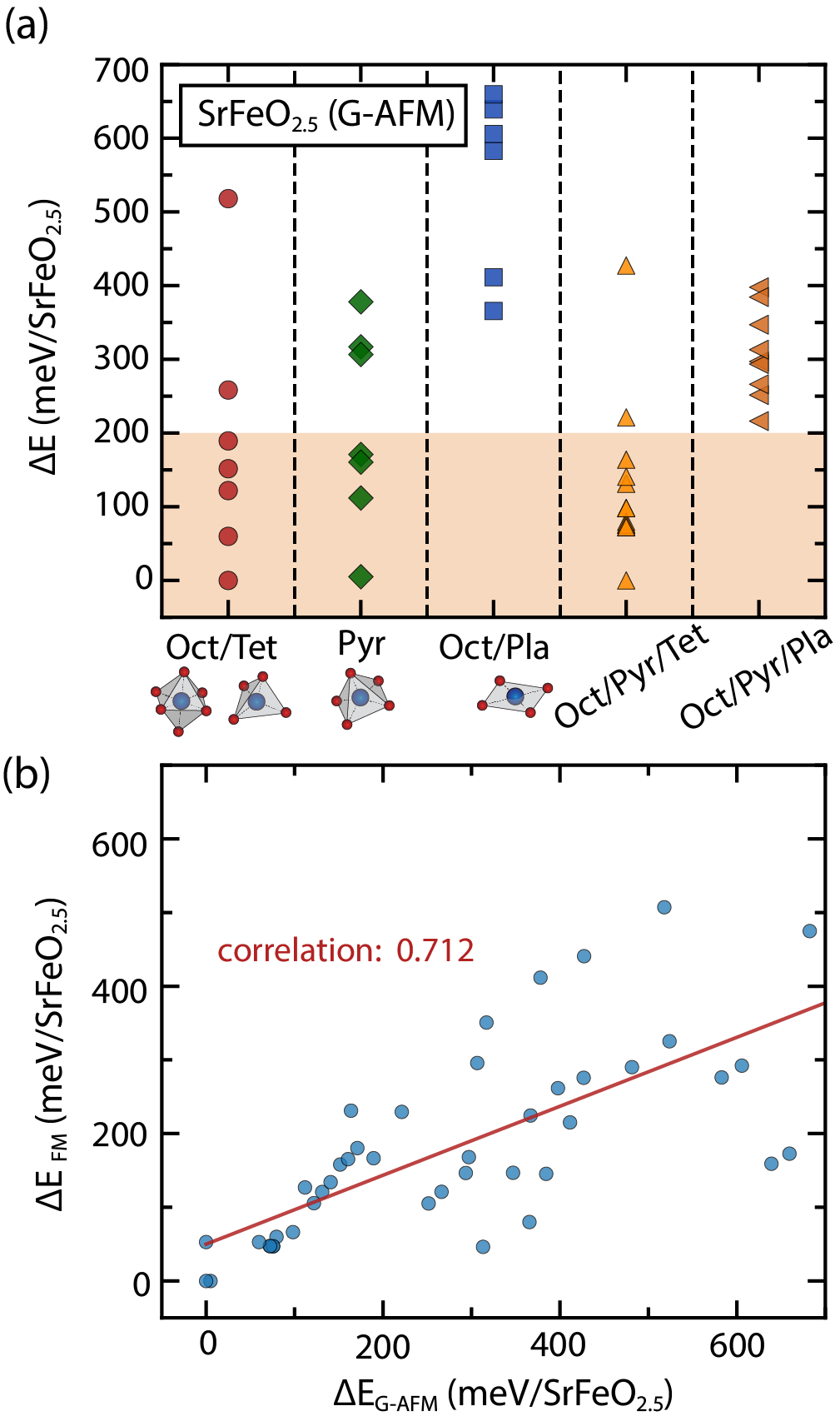}
\caption{(a) DFT-level calculated energies of SrFeO$_{2.5}$ OOV phases, classified by constituent FBUs. G-type antiferromagnetic (G-AFM) order is assumed. (b) Correlation between energies with ferromagnetic and G-AFM order exhibit a Pearson correlation of 0.712. 
}
\label{fig:SFO_energies}
\end{figure}

\subsubsection{Magnetic interactions}

In \autoref{fig:SFO_energies}(a),
we present the energy landscape of \SFO as a magnetic analogue to \CAO. Both compounds adopt the brownmillerite structure in their ground state, which allows us to study the contribution of magnetic interactions to the OOV stability.
The energies of \SFO were calculated with G-type antiferromagnetic (G-AFM) order as predicted by the GKA-rule based on a $d^5$ electronic configuration of Fe$^{3+}$ in \SFO.
When plotting the energy with respect to the FBU groups for \SFO, we find a trend similar to that found for \CAO.
The square planar units are energetically unfavored.

We examined the magnetic contributions by comparing the correlation between DFT-level energies between ferromagnetic (FM) and G-AFM order in \autoref{fig:SFO_energies}(b), where atomic relaxations are also performed separately.
As the five half-filled $3d$-orbital of Fe$^{3+}$ yields a local magnetic moment of $\sim$5$\mu_B$, 
FM order in \SFO is strongly disfavored and should be the most unstable magnetic configuration.
The correlation between G-AFM and FM order serves as a proxy for the upper-limit of the energy variations  created by the magnetic interactions.
Overall, the change in the magnetic order leads to dispersion in the energies for each structure and a  correlation of 0.712.
The dispersion is large for relatively unstable OOV phases ($>200$\,meV/\SFO), whereas the low-energy phases ($<200$\,meV/\SFO) show a high correlation (0.934) between the two magnetic settings.
This relation indicates that the stability of the OOV structures is more dominated by the FBU-assemblies and relevant elastic costs, while the magnetic interaction play secondary roles.
Thus, we anticipate FM ordering to be a suitable model to search for stable OOV phases in magnetic OOV oxides.

\subsection{Design of new OOV compounds}
\label{ss:design}

\subsubsection{Structure-type design}

\begin{figure}[t]
\centering
\includegraphics[width=0.93\columnwidth]{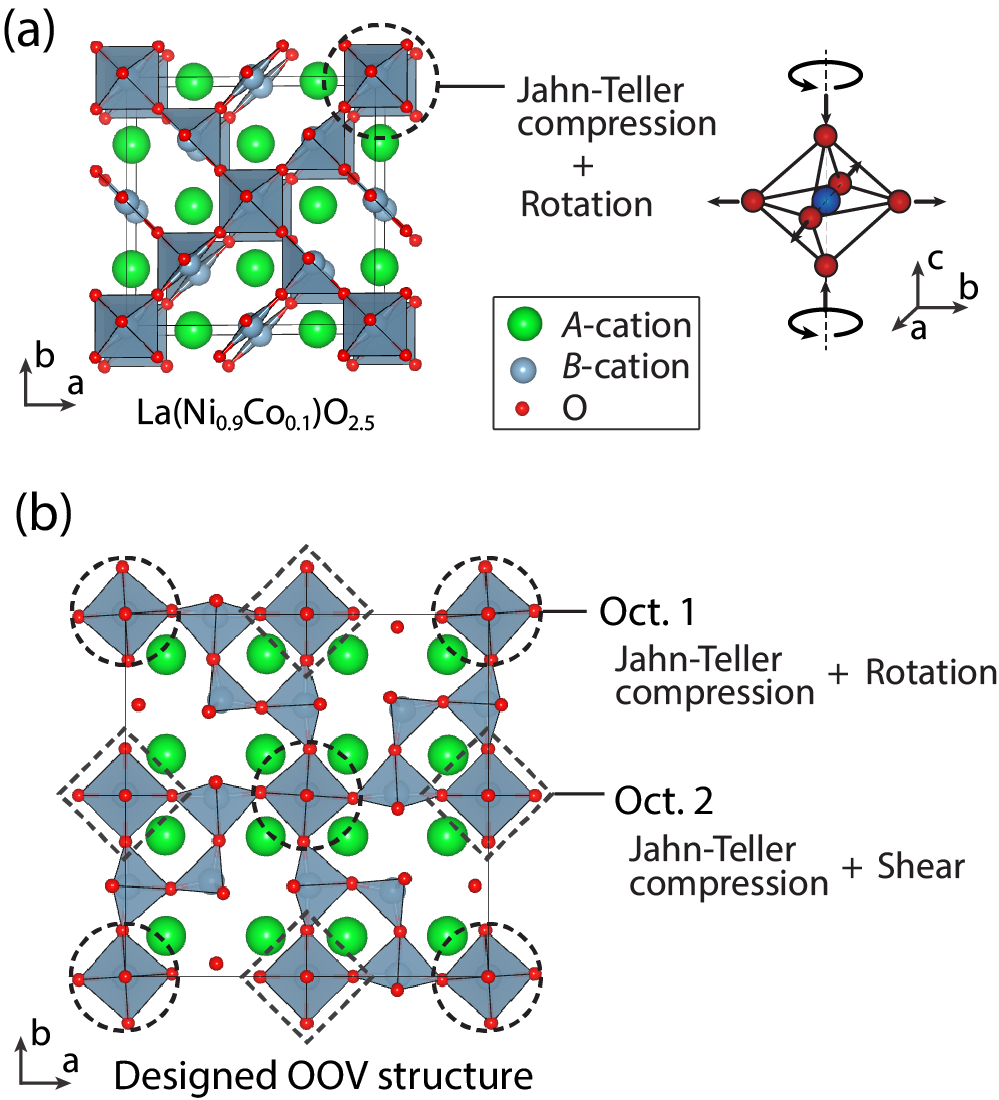}
\caption{Atomic structures of OOV phases with greater periodicity in the (111)$_\mathrm{pc}$ plane than that found in our 50-structure data set. (a) La(Ni$_{0.9}$Co$_{0.1}$)O$_{2.5}$ structure with octahedra, pyramids, and square planes. (b) Designed OOV structure with octahedra, pyramids, and tetrahedral units. Two symmetrically different octahedra are highlighted with empty dashed circle (Oct.\ 1) and dashed square (Oct.\ 2).
}
\label{fig:new_structures}
\end{figure}

Summarizing the ordering principles, we identified the following criteria for stable OOV structures: (i) chemical preference of transition metals to the FBU-type, (ii) column-wise aligned polyhedral units with vacancy channels, and (iii) assembly allowing for distinct bond lengths/angles of FBUs with low-energy distortions.
We found these criteria can be generalized to larger unit cells beyond those obtained from our generated structures.
For example, the structure of \autoref{fig:new_structures}(a) composed with square planar, square pyramids, and octahedral units, is realized in nickel- or copper-based compounds with mixed $A$ or $B$ cations \cite{Aasland1998,Genouel/Raveau1995,Er-Rakho/Raveau1988}.
In this structure, the multiple elements and charge states stabilize three different types of FBUs. Each FBU-type is parallelly aligned in columnar arrangement with vacancy channels along [100]\PC. 
Additionally, the distinct bond lengths of constituent FBUs (including different apical/basal bonds in pyramids), are assembled with mainly rotations and Jahn-Teller like distortions. As discussed in our use cases, these distortions cost minimal energy and this structure satisfies the criteria stated above.
Indeed, the calculated DFT energy of this structure in CaAlO$_{2.5}$ chemistry is 402\,meV/\CAO in  \autoref{fig:CAO_energies}(a), this value is close to the minimum energy in the category of the same Oct/Pyr/Pla group.

In \autoref{fig:new_structures}(b), we designed a new OOV structure by  substituting the planar units in \autoref{fig:new_structures}(a) with tetrahedra and additional octahedra and pyramids.
The octahedron at the center of the structure (Oct.\,1) has four neighboring pyramids arranged like a windmill, similar to the octahedra in \autoref{fig:new_structures}(a).
In this designed OOV structure, there is another octahedral unit with different site symmetry (Oct.\,2 in \autoref{fig:new_structures}). In Oct.\,2 two out of four neighboring pyramids are connected to the same tetrahedron.
Owing to the difference in the connectivity, these two octahedra are subjected to different types of distortions: Oct.\,1 is rotated and Oct.\,2 is sheared.
As mentioned in the FBU-assemblies use cases, the shearing mode is not considered as a low-energy distortion.
Indeed, we find the calculated $\Delta$E of this structure for \CAO is 199.4\,meV/\CAO referenced to the lowest energy in Oct/Pyr/Tet group. This energy value falls within the middle of the range of energies calculated for other structures in the Oct/Pyr/Tet group 
\autoref{fig:new_structures}(a). 
Considering that the shearing mode leads to higher energy for \CAO,
we anticipate other chemistries may further stabilize this new OOV structure by suppressing the shearing distortion.
Specifically, \ABO chemistries with tolerance factors close to unity will have oxygen coordinates close those in cubic perovskite, reducing the shearing motion of the octahedral units. %

\subsubsection{Chemical selection} 

\begin{figure}[t]
\centering
\includegraphics[width=0.75\columnwidth]{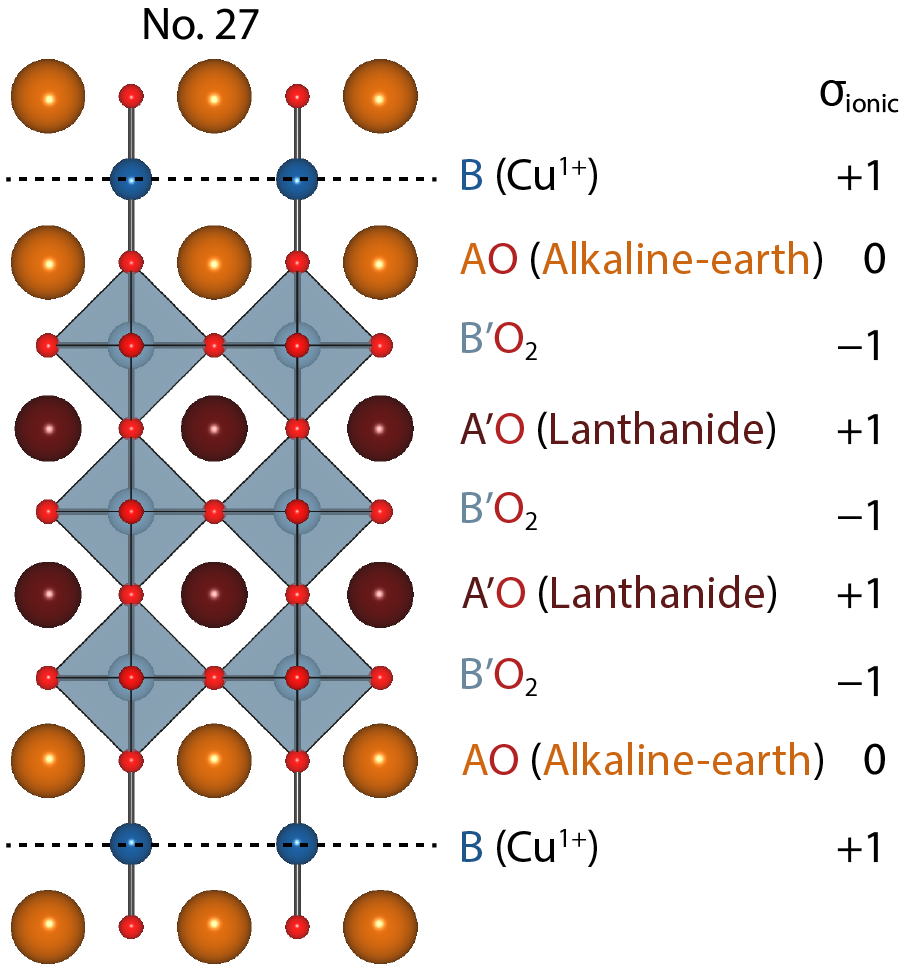}
\caption{Atomic structures of No.\,27 with assumed chemistry and cation order. Layer charge densities are  based on ionic charges in a given layer as specified on the right.}
\label{fig:Design27}
\end{figure}

Next, we demonstrate an examplary design of a new chemistry with an OOV pattern that is yet to be synthesized.
We focus on structure No.\,27 in \autoref{fig:Design27}, consisting of three octahedral layers connected by a layer of linear units ($B$O$_2$).
We begin the chemical selection process by searching for transition metals in charge states that will occupy the non-octahedral units.
A well-known 2-coordinate cation is Cu$^{1+}$, so we select it for the linear coordination sites. 
Then we need to stablize this charge state by choice of cations for the octahedral $B^\prime$ and $A$ sites. 

The simplest approach might be to construct the entire structure only with Cu with the selection of appropriate $A$-cations, as Cu$^{1+}$, Cu$^{2+}$, and Cu$^{3+}$ can occupy the octahedral sites. 
However, this approach is not realistic given the interactions and features stablizing OOV phases. 
Cu$^{1+}$ on the $B$ site of $AB$O$_{2.5}$ restricts the charge state of the $A$-cation to 4+ so as to satisfy charge neutrality ($A^{4+}$Cu$^{1+}$O$_{2.5}^{2-}$); thus, exclusive occupancy of the $A$ site by alkali, alkaline-earth metals and lanthanides is not possible.
In addition, because some $A$O layers in the structure are between linear and octahedral layers, selection of the $B$ cation and its charge state should be accommodated by splitting the $A$ cation site and occupying it by cations with 1+ and 3+ oxidation states.
One option is to all Cu to take different charge states in the octahedral units, such as 
Cu$^{2+}$(Oct) or Cu$^{3+}$(Oct).
The scenario with Cu$^{2+}$(oct) would require one of the $A$ cations to be quatravalent, so we do not consider this case further. 
Cu$^{3+}$(Oct) is compatible with more common divalent and trivalent $A$ cations (two  $A^{2+}$ and two $A^{3+}$) found in perovskites; 
however, the charge disproportionation requires a strong driving force to overcome the stability of a uniform charge distribution with homogeneous Cu$^{2+}$ cations on all sites.
The competing phases involve various polyhedral combination like Oct/Tet and Oct/Pla groups.
Given the more extreme disproportionation in both charge states (Cu$^{+}$ and Cu$^{3+}$) and coordination (linear unit and octahedra),
the stability of structure No.\,27 is not guaranteed.

Given the unlikelihood of Cu being the only $B$ cation in structure No.\,27 to stablize the compound, we examine whether introducing other $B$ cations ($B'$) in the octahedral units would improve the stability of the structure.
From the discussion above, 
the desired feature of $B'$ is to present stablity exclusively in octahedral units with a charge state equal to or larger than +3.
The candidates satisfying this condition include Fe$^{3+}$, Al$^{3+}$, Sc$^{3+}$, Y$^{3+}$, In$^{3+}$, and Cr$^{3+}$.
To maintain the low charge state of Cu$^{1+}$,
the redox potential of $B'^{3+}$ needs to be smaller than that of Cu$^{1+} \rightleftharpoons$ Cu$^{2+}$ + $e$ (0.153\,V). 
\autoref{table:redox} provides the redox potentials for these candidates cations.
The redox potential of Fe$^{3+}$ is 0.771\,V, meaning that the combination of  Cu$^{1+}$ and Fe$^{3+}$ is less stable than
Fe$^{2+}$ and Cu$^{2+}$. Thus Fe is excluded from the candidates.

\begin{table}[t]
\caption{\label{table:redox} 
Standard electrode potentials ($E^\circ$)\cite{CRChandbook} of candidate $B'$ cations for structure No.\,27, and ionic radii of $A$ cations with different values of Goldschmidt tolerance factor \cite{Shannon1976,Goldschmidt1926}.
}
\centering
\begin{tabular}{l r r  c c c }
\toprule
     Element   &    \multicolumn{2}{c}{Redox potential}            &     \multicolumn{3}{c}{Ionic radius of $A$ (\AA)} \Bstrut \\
         \cline{2-3} \cline{4-6}
         \Tstrut
                   
                    &Reaction & $E^\circ/V$  & $t=$ 0.83 & $t=$ 1  &  $t=$ 1.06 \Bstrut \\ 
\midrule
Cu$^{1+}$     & oxidation   &  $-$0.153 &   0.76    &   1.21    & 1.36   \\     
Al$^{3+}$     & reduction  & $-$1.662    &   0.85    &   1.32    & 1.47   \\
Cr$^{3+}$    &  reduction &  $-$0.407    &   0.94 	&    1.43   & 1.59   \\
Fe$^{3+}$    & reduction  &    0.771      &   0.98    &   1.47   & 1.64    \\
Sc$^{3+}$    & reduction  &$-$2.077     &   1.09    &    1.61   & 1.79   \\
In$^{3+}$      & reduction & $-$0.490     &   1.16    &   1.69   & 1.87    \\
Y$^{3+}$      & reduction  &  $-$2.372   &   1.28    &   1.83   & 2.02    \\
\bottomrule
\end{tabular}
\end{table}

As the possible $B'$ cations are specified, now we search for appropriate $A$ cations.
The $A$ cations needs to give an average +2.5 charge to satisfy neutrality in the formula $A^{2.5+}_4$Cu$^{1+}$$B'$$_3^{3+}$O$_{10}^{2-}$.
Thus, splitting the $A$ site and occupying it with an alkaline-earth metal ($A^{2+}$) and lanthanides ($A'^{3+}$) should stabilize the structure.
Before discussing whether these cations possess suitable ionic radii, 
we assume the structure will adopt $A$/$A'$ cation order as depicted in  \autoref{fig:Design27}.
The reasoning behind this assumption is based on the electrostatic potential profile induced by cation ordering of the inequivalent charge states from the $A$ and $A'$ cations \cite{Shin2017}.
The Cu$^+$ layer has nominal charge of +1 (unit of $e$/$a_p^2$), and the neighboring atomic layers is $A$O with a charge of 0 (neutral) if the $A$ cation is an alkaline-earth metal, +1 if the $A$ cation is a  lanthanide, and +0.5 if the two $A$-cations are mixed.
The internal electric field obtained by integrating the layer charge densities, so the successive positively charged layers (which includes negative charge-dominant octahedral regions in the structure) would form an extremely high internal electric field \cite{Shin2017}.
Thus, having lanthanides near the Cu$^+$ layer would not be a feasible solution.

Although the electrostatic instability needs to be considered as a significant contribution,
we note that this contribution can be overcome by cation size effects in OOV structures.
In (Ho,Ba)FeO$_{2.5}$ with structure No.\,1 in  \autoref{generated_structures}(c),\cite{Woodward2003} the lanthanide ion (Ho) is positioned between the basal planes of square pyramids, while Ba is at the $A$O layer of the apical oxygen.
This configuration would lead to a higher electrostatic instability than the opposite arrangement (having Ho at the $A$O layer and Ba between the basal planes).
In this case. the ionic radius of Ho is considerably smaller than that of Ba, so Ho fits better in the narrower space provided by the ordered oxygen vacancy plane, optimizing the bond valence of the constituent cations.
In addition, although the nominal charge of the $A$ layer is +3, the neighboring layers is FeO$_2$ with a nominal charge of -1.5, where the relatively high positive charge of the $A$ layer is partly canceled within a small distance.

The last consideration for chemical selection of elements that would be stable in structure No.\,27 involves the ionic radii of the $A$ and $B$ cations.
When choosing the lanthanide element ($A'$), the Goldshmidt tolerance factor\cite{Goldschmidt1926} can be used, because the majority of the structure comprises perovskite blocks.
The stable tolerance factor range for perovskites is between 0.825 and 1.059 \cite{Barteleaav2019}.
Among $B$-cation candidates, Al and Y exhibit the smallest and largest  ionic radii (0.535 and 0.9\,\AA ), and the corresponding $A$-cation range is from 0.85 to 1.47\,\AA\ and from 1.275 to 2.02\,\AA, respectively.
Thus, La fits the constraints for all $B$-cation candidates, but its combination with Y yields a tolerance factor of 0.85. This low tolerance factor should produce considerable octahedral rotations that would inhibit the formation of the OOV structure.

Similarly, the Goldschmidt tolerance factor indicates that relatively smaller alkaline-earth cations ($A$) could be stable in the structure with some potential variability.
If the same criterion is applied, then Ca is the only appropriate sized $A$ cation (radius of 1.34\,\AA\xspace as a 12-coordinated cation).
Since the Goldschmidt tolerance factor assumes close packed of ions and the $A$ cations near Cu$^{1+}$ exhibit 8 coordination due to the oxygen vacancies, we expect larger $A$ cations than would nominally be indicated from tolerance factor could be more stable to compensate the decreased bond valence resulting from the OOVs.
Therefore, Sr (1.44\,\AA) or Ba (1.61\,\AA) could also be selected.
We note that the alkaline-earth cations connect the linear and octahedral units,
so its relative size to the $B'$-cations can also guide selection to assess stability of the OOV phases.

With these guidelines, one can than perform ab initio thermodynamic simulations to assess the phase stability of the candidate compounds with structure No.\,27 
in the family (Sr,Ba)$_{0.5}$La$_{0.5}$Cu$_{0.25}$$B'_{0.75}$O$_{2.5}$ ($B'$ = Al, Cr, Sc, In, Y).
To summarize the design approach, we followed the steps:
(i) compatible $B$ cation for non-octahedral unit (Cu$^+$),
(ii) possibility of disproportionation in charge states or polyhedral unit (Cu$^{2+/3+}$ in octahedral unit),
(iii) redox potential of the other polyhedral units (exclusion of Fe$^{3+}$),
(iv) required charge states of $A$ cations (combination of alkali/alkaline-earth/lanthanides),
(v)  potential electrostatic instability derived from cation orders ($A$/$A'$ sites are designated),
(vi) ionic radii within the stable tolerance factor range, adjusted with actual valence of each $A$ sites.
We expect this chemistry-selection procedure can be applied to any OOV phases beyond $AB$O$_{2.5}$ stoichiometry.


\section{Conclusion}
We examined the stability of OOV phases by using an efficient algorithm for constructing oxygen-deficient perovskite structures with ordered oxygen vacancies (OOVs), focusing on $AB$O$_{2.5}$ stoichiometry.
By utilizing the experimental observation of an homogeneous vacancy patterns on the (111)$_\mathrm{pc}$ planes, we efficiently enumerated a comprehensive set of OOV phases.
The structure set enabled us to examine the major contributions to stability of the OOV patterns.
We identified that the primary criteria determining the stability includes the preference of elements toward different fundamental building unit types, and allowed atomic relaxations in the given vacancy arrangements.
Specifically, the advantage of forming vacancy channels is essential to the stability. 
The vacancy channels are often compatible with the cooperative motions of close-packed layers in the structures, which enables better optimization of the bond valences of the comprising cations.
%
Our analysis explains the known trends in exisiting OOV structures such as coordination number and ionic sizes of $A$ and $B$ cations, and the
ordering principles gleaned from our analysis can be used to design novel OOV structures.
Lastly, we enumerated and employed chemical selection criteria to find a family of oxides that should be compatible with a given OOV structure.
This design procedure can be applied to other OOVs phases beyond those specifically considered here. 
Our study on OOVs contributes better understanding of the materials chemistry aspects involved in oxides with ordered oxygen vacancies, and may be used to 
discover metastable and 
nonequilibrium bulk or thin film oxides.

\begin{acknowledgements}
This work was supported in part by SUPREME, one of seven centers in JUMP 2.0, a Semiconductor Research Corporation (SRC) program sponsored by DARPA. 
K.R.P.\ was supported by the National Science Foundation (DMR-1904701).
Calculations were performed using the QUEST HPC Facility at Northwestern, the Extreme Science and Engineering Discovery Environment (XSEDE), which is supported by the National Science Foundation under Grant No.\ ACI-1548562, and the Center for Nanoscale Materials (Carbon) Cluster, an Office of Science user facility supported by the U.S.\ Department of Energy, Office of Science, Office of Basic Energy Sciences, under Contract No.\ DE-AC02-06CH11357.
\end{acknowledgements}

\section*{Supplementary Information}
The Supporting Information is available free of charge on the
publisher's website at DOI: [\textit{\textbf{add doi and website link}}].
\begin{enumerate}
    \item[] %
	2D planar vacancy patterns observed in $AB$O$_{2.5}$ compounds; 
	crystallographic information for OOV structures; energies of LaNiO$_{2.5}$ OOV phases; analysis of atomic
	relaxations of square pyramidal networks; correlation of structural descriptors and energetics of 
	square pyramidal networks; and atomic displacements of CaAlO$_{2.5}$ structures in the Oct/Tet group 
	and Oct/Pla group.	
\end{enumerate}

\bibliographystyle{apsrev4-1}
\bibliography{YShin}

\newpage\newpage



\section*{Supplementary material}

# Informatics-based learning of oxygen vacancy ordering principles in oxygen-deficient perovskites

Yongjin Shin[1], Kenneth R. Poeppelmeier[2], and James M. Rondinelli[3]

[1]*Pritzker School of Molecular Engineering, University of Chicago, IL 60637, USA*
[2]*Department of Chemistry, Northwestern University, Evanston, IL 60208, USA*
[3]*Department of Materials Science and Engineering, Northwestern University, Evanston, IL 60208, USA*

**Contents**

# I. Surveyed planar patterns on (111)$_{pc}$ layer

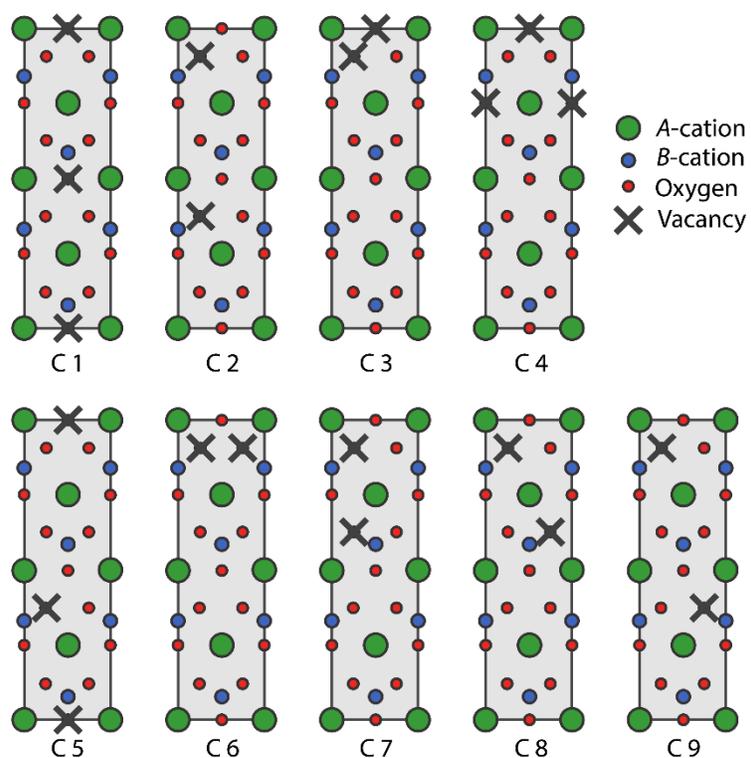

**Figure S1 |** Nine unique planar vacancy configurations on (111)$_{pc}$ layer constituting $ABO_{2.5}$ stoichiometry, labeled C1 to C9. Large Green, small blue, and small red circles represent the $A$ cations, $B$ cations, and oxide anions, while black crosses indicate oxygen vacancies. We note oxygen vacancies are defined on $AO_3$ layers whereas $B$ cations occupy octahedral interstitial sites between the (111)$_{pc}$ layers.

## II. List of generated OOV structures

**Table S1** | Generated structures with space groups and polyhedral configurations. Polyhedral configurations are described by FBU type: Octahedral (Oct), Square pyramidal (Pyr), Tetrahedral (Tet), Square planar (Pla), Triangular planar (TriPlane), Tri-legged (TriLeg), and Linear (Line). $(111)_{pc}$ pattern corresponds to labels in Figure S1, and translational vector ($\vec{t}$) follows the vectors shown in Fig. 2 in the main text. The rightmost column indicates whether the handedness of the layer pattern alters with stacking of $(111)_{pc}$ layers.

| No. | Space group | SG# | Polyhedral config. | $(111)_{pc}$ pattern | $\vec{t}$ | Handedness change with stacking |
|-----|-------------|-----|---------------------|---------------------|-----------|----------------------------------|
| 1 | *P 4/m m m* | 123 | Pyr, Pyr, Pyr, Pyr | C1 | $\vec{t}_1$ | X |
| 2 | *P 4/m m m* | 123 | Oct, Oct, Pla, Pla | C1 | $\vec{t}_2$ | X |
| 3 | *P m m m* | 47 | Oct, Oct, Pla, Pla | C2 | $\vec{t}_1$ | X |
| 4 | *P m m a* | 51 | Oct, Oct, Tet, Tet | C2 | $\vec{t}_1$ | O |
| 5 | *C m m m* | 65 | Pyr, Pyr, Pyr, Pyr | C2 | $\vec{t}_2$ | X |
| 6 | *P m n a* | 53 | Pyr, Pyr, Pyr, Pyr | C2 | $\vec{t}_2$ | O |
| 7 | *P m m 2* | 25 | Oct, Oct, Pyr, TriPlane | C3 | $\vec{t}_1$ | X |
| 8 | *P m a 2* | 28 | Oct, Oct, Pyr, TriLeg | C3 | $\vec{t}_1$ | O |
| 9 | *P 1* | 1 | Oct, Pyr, Pyr, Tet | C3 | $\vec{t}_2$ | X |
| 10 | *P c* | 7 | Oct, Pyr, Pyr, Tet | C3 | $\vec{t}_2$ | O |
| 11 | *C m* | 8 | Oct, Pyr, Pyr, Tet | C3 | $\vec{t}_3$ | X |
| 12 | *P c* | 7 | Oct, Pyr, Pyr, Tet | C3 | $\vec{t}_3$ | O |
| 13 | *P m* | 6 | Oct, Oct, Pyr, TriPlane | C3 | $\vec{t}_4$ | X |
| 14 | *P m a 2* | 28 | Oct, Oct, Pyr, TriPlane | C3 | $\vec{t}_4$ | O |
| 15 | *P 4/m m m* | 123 | Oct, Pyr, Pyr, Pla | C4 | $\vec{t}_1$ | X |
| 16 | *C m c m* | 63 | Pyr, Pyr, Pyr, Pyr | C4 | $\vec{t}_2$ | X |
| 17 | *I 4/m m m* | 139 | Oct, Pyr, Pyr, Pla | C4 | $\vec{t}_3$ | X |
| 18 | *P m m a* | 51 | Oct, Oct, Pla, Pla | C4 | $\vec{t}_4$ | X |
| 19 | *P m m 2* | 25 | Oct, Pyr, Pyr, Pla | C5 | $\vec{t}_1$ | X |
| 20 | *P m a 2* | 28 | Oct, Pyr, Pyr, Tet | C5 | $\vec{t}_1$ | O |
| 21 | *P 1* | 1 | Oct, Pyr, Pyr, Tet | C5 | $\vec{t}_2$ | X |
| 22 | *P c* | 7 | Oct, Pyr, Pyr, Tet | C5 | $\vec{t}_2$ | O |
| 23 | *C m* | 8 | Oct, Pyr, Pyr, Tet | C5 | $\vec{t}_3$ | X |
| 24 | *P c* | 7 | Oct, Pyr, Pyr, Tet | C5 | $\vec{t}_3$ | O |
| 25 | *P m* | 6 | Oct, Pyr, Pyr, Pla | C5 | $\vec{t}_4$ | X |
| 26 | *P m a 2* | 28 | Oct, Pyr, Pyr, Pla | C5 | $\vec{t}_4$ | O |
| 27 | *P 4/m m m* | 123 | Oct, Oct, Oct, Line | C6 | $\vec{t}_1$ | X |
| 28 | *C m c m* | 63 | Oct, Oct, tet, Tet | C6 | $\vec{t}_2$ | X |
| 29 | *I m m a* | 74 | Oct, Oct, tet, Tet | C6 | $\vec{t}_3$ | X |
| 30 | *P m m a* | 51 | Oct, Oct, tet, Tet | C6 | $\vec{t}_4$ | X |
| 31 | *P -4 m 2* | 115 | Oct, Oct, Pla, Pla | C7 | $\vec{t}_1$ | X |

| 32 | *P m m a* | 51 | Oct, Oct, tet, Tet | C7 | $\vec{t}_1$ | O |
| 33 | *C m m 2* | 35 | Oct, Pyr, Pyr, Tet | C7 | $\vec{t}_2$ | X |
| 34 | *P m n a* | 53 | Oct, Pyr, Pyr, Pla | C7 | $\vec{t}_2$ | O |
| 35 | *I 4₁ 2 2* | 98 | Pyr, Pyr, Pyr, Pyr | C7 | $\vec{t}_3$ | X |
| 36 | *P m n a* | 53 | Pyr, Pyr, Pyr, Pyr | C7 | $\vec{t}_3$ | O |
| 37 | *P m m 2* | 25 | Oct, Pyr, Pyr, Tet | C7 | $\vec{t}_4$ | X |
| 38 | *P c c m* | 49 | Oct, Pyr, Pyr, Pla | C7 | $\vec{t}_4$ | O |
| 39 | *P m m m* | 47 | Oct, Oct, Pla, Pla | C8 | $\vec{t}_1$ | X |
| 40 | *P m n a* | 53 | Oct, Oct, Tet, Tet | C8 | $\vec{t}_1$ | O |
| 41 | *C 2/m* | 12 | Oct, Pyr, Pyr, Pla | C8 | $\vec{t}_2$ | X |
| 42 | *P m m n* | 59 | Oct, Pyr, Pyr, Tet | C8 | $\vec{t}_2$ | O |
| 43 | *F m m m* | 69 | Pyr, Pyr, Pyr, Pyr | C8 | $\vec{t}_3$ | X |
| 44 | *P n n a* | 52 | Pyr, Pyr, Pyr, Pyr | C8 | $\vec{t}_3$ | O |
| 45 | *P 2/m* | 10 | Oct, Pyr, Pyr, Pla | C8 | $\vec{t}_4$ | X |
| 46 | *P m m a* | 51 | Oct, Pyr, Pyr, Tet | C8 | $\vec{t}_4$ | O |
| 47 | *P 4₂/m m c* | 131 | Oct, Pyr, Pyr, Pla | C9 | $\vec{t}_1$ | X |
| 48 | *C m c a* | 64 | Pyr, Pyr, Pyr, Pyr | C9 | $\vec{t}_2$ | X |
| 49 | *I m m a* | 74 | Oct, Oct, Tet, Tet | C9 | $\vec{t}_3$ | X |
| 50 | *P b a m* | 55 | Pyr, Pyr, Pyr, Pyr | C9 | $\vec{t}_4$ | X |

### III. Energy landscape of LaNiO$_{2.5}$ OOV phases

Fig. S2 shows the clustering of energy values of LaNiO$_{2.5}$ for different groups of polyhedral units in the OOV phase. Non-spin polarized DFT was used to assess the OOV phases. In contrast to the energy landscape of CaAlO$_{2.5}$ shown in the main text, we find notable preference of Ni$^{2+}$ toward square planar units. The lowest energy was found within the Oct/Pyr/Pla group, whose order was previously observed in La(Ni$_{0.9}$Co$_{0.1}$)O$_{2.5}$ [1]. The structure with lowest energy in the Pyr group is Ca$_2$Mn$_2$O$_5$-type structure (No. 50) and its energy was 25.58 meV/f.u. higher than that of La(Ni$_{0.9}$Co$_{0.1}$)O$_{2.5}$ structure. Within the Oct/Pla group, we found the energy of No.2 (previously known LaNiO$_{2.5}$ structure) is 49.2 meV/f.u. higher than No. 3. As mentioned in the main text, the structures were relaxed with the initial symmetry imposed by the OOV patterns. The square planar units preserve relatively higher symmetry than the other structures, without showing rotation/tilting distortions. Thus, we expect the energies of structures with planar group would be further lowered if additional distortions were permitted.

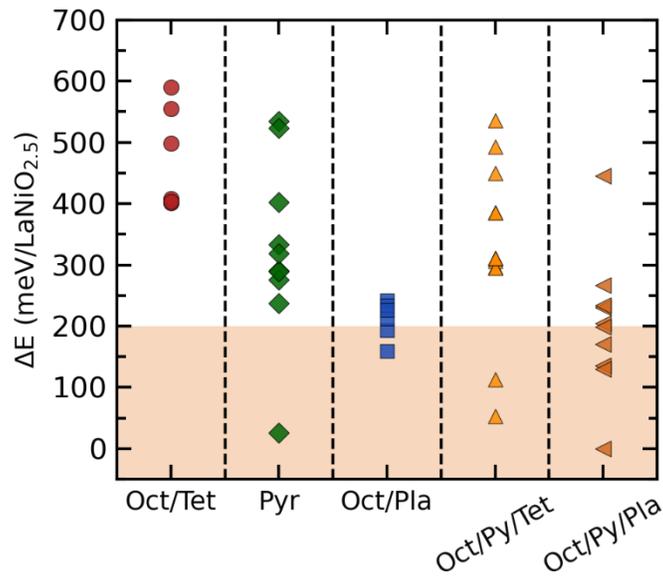

**Figure S2** | Calculated energies of LaNiO$_{2.5}$ OOV phases with non-spin-polarized DFT, classified by constituent functional building units. Each structure is relaxed within the initial symmetry imposed by OOVs.

## IV. Atomic relaxation of square pyramidal network (No. 50)

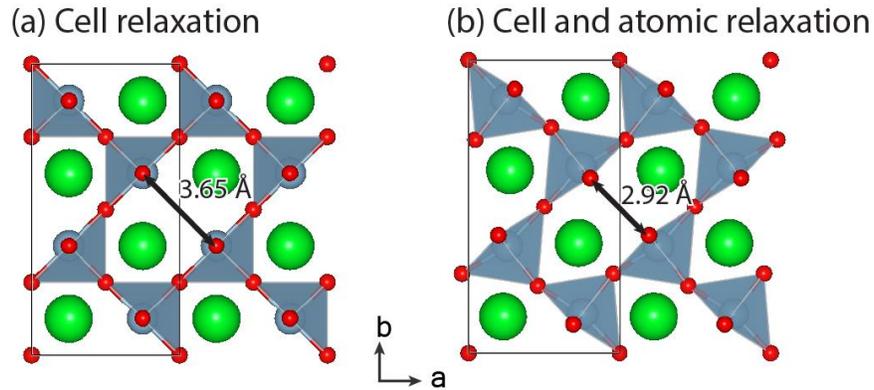

(a) Cell relaxation        (b) Cell and atomic relaxation

3.65 Å

2.92 Å

**Figure S3** | Comparison of atomic coordinates in Ca$_2$Mn$_2$O$_5$-type structure (No. 50) with (a) cell relaxation and (b) cell and atomic relaxation setting. The basal oxygen atoms displace toward the vacancy channel, while Al atoms displace into the pyramid. The resultant distance between two oxygen atoms across the vacancy channel decreases from 3.65 Å to 2.92 Å. Ca, Al, and O atoms are represented by large green, blue, and small red spheres, respectively.

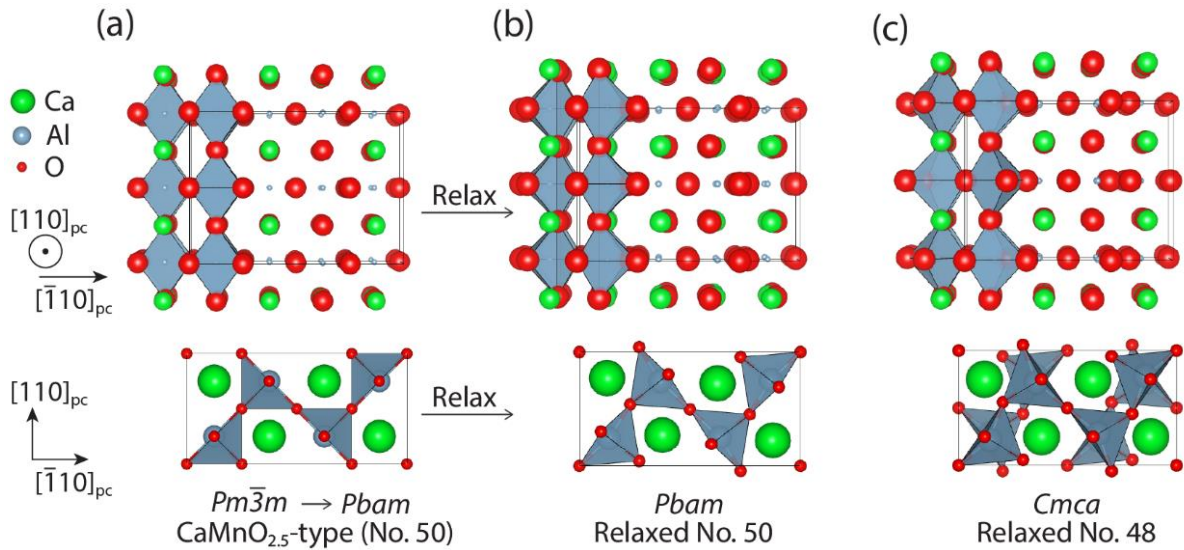

(a)        (b)        (c)

Ca
Al
O

[110]$_{pc}$

[$\bar{1}$10]$_{pc}$

[110]$_{pc}$

[$\bar{1}$10]$_{pc}$

Relax

Relax

$Pm\bar{3}m \rightarrow Pbam$
CaMnO$_{2.5}$-type (No. 50)

$Pbam$
Relaxed No. 50

$Cmca$
Relaxed No. 48

**Figure S4** | Atomic relaxation of CaAlO$_{2.5}$ with square pyramidal network, No. 50 and No. 48, viewed from the [110]$_{pc}$ direction. (a) Structure No. 50 generated from cubic perovskite without atomic relaxation. All AO and OO chains are linear without cooperative displacements. (b) Atomic relaxation of structure No. 50 involves displacements of all AO chains and half of the OO chains along the [$\bar{1}$10]$_{pc}$ direction. (c) Atomic relaxation of structure No. 48 mainly results in half of the OO chains displacing while the AO chains are inactive. Ca, Al, and O atoms are represented by large green, blue, and small red spheres, respectively.

**V. Correlation of structural descriptors and energetics of square pyramidal networks**

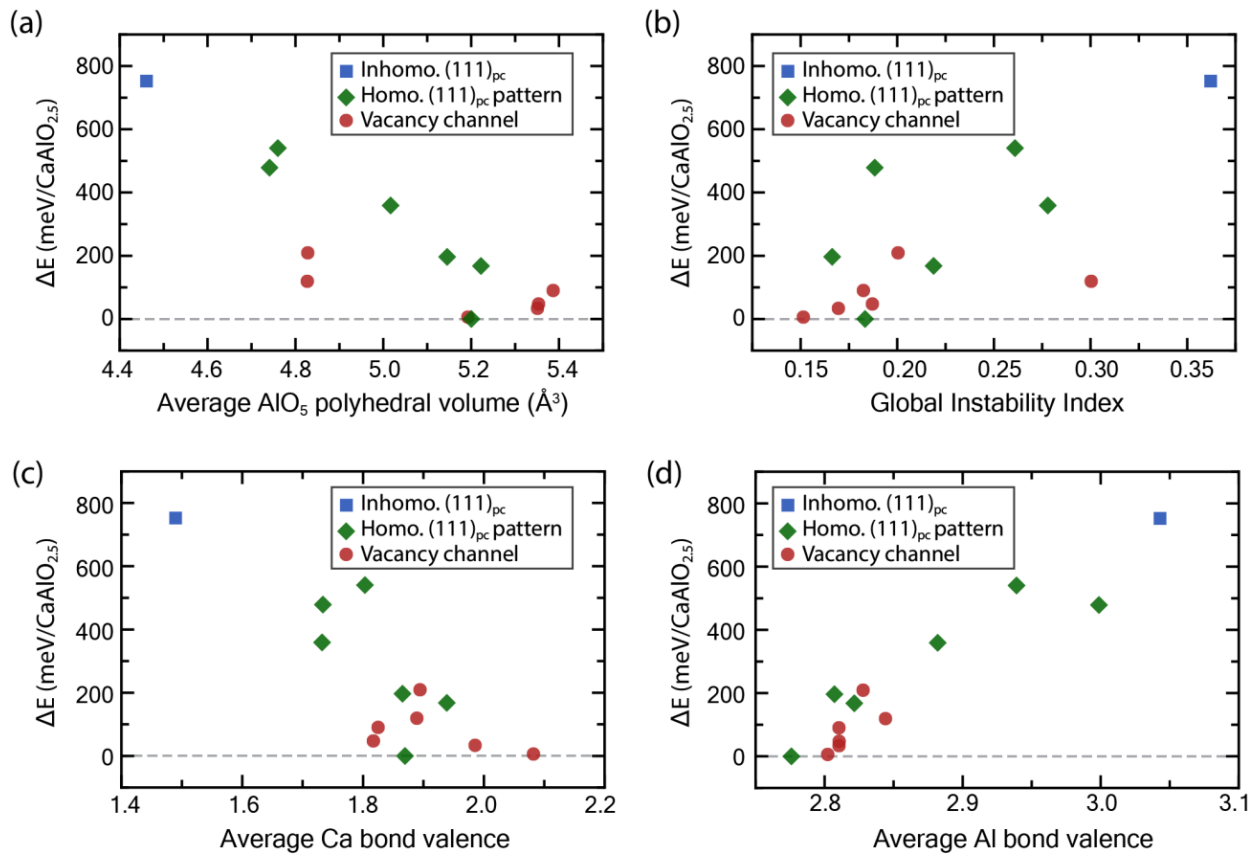

**Figure S5** | Energy differences of CaAlO$_{2.5}$ square pyramidal networks plotted as a function of various structural descriptors: (a) Average polyhedral volume of the AlO$_5$ square pyramid, (b) global instability index, (c) average Ca bond valence, and (d) average Al bond valence.

**VI. Atomic displacements of CaAlO$_{2.5}$ structures in Oct/Tet group**

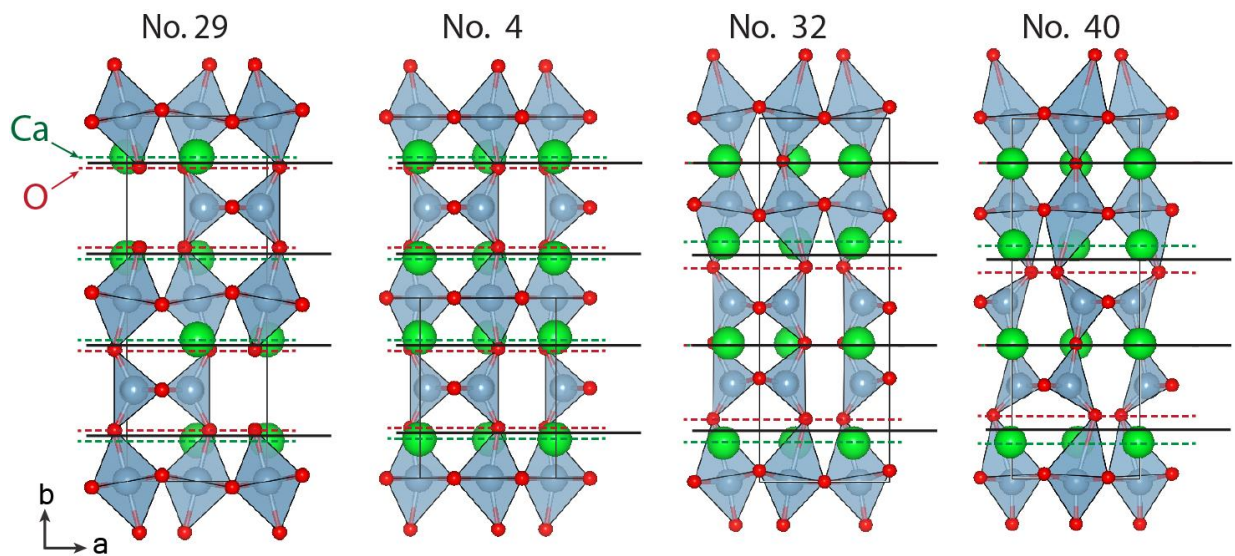

**Figure S6** | Atomic relaxation of selected Oct/Tet structures with CaAlO$_{2.5}$ chemistry. Compared to the coordinates of cubic perovskites, the Ca and O atoms displace toward opposite directions. This displacement leads to a splitting in the *b* coordinates of Ca and O atoms, as indicated by dashed green and red lines. This displacement is captured by propagation of a distortion in the CaO chain, where the center of Ca and O atoms (solid black line) is regularly spaced in perovskite lattice.

## VII. Atomic relaxation of perturbed CaAlO₂.₅ structures in Oct/Pla group

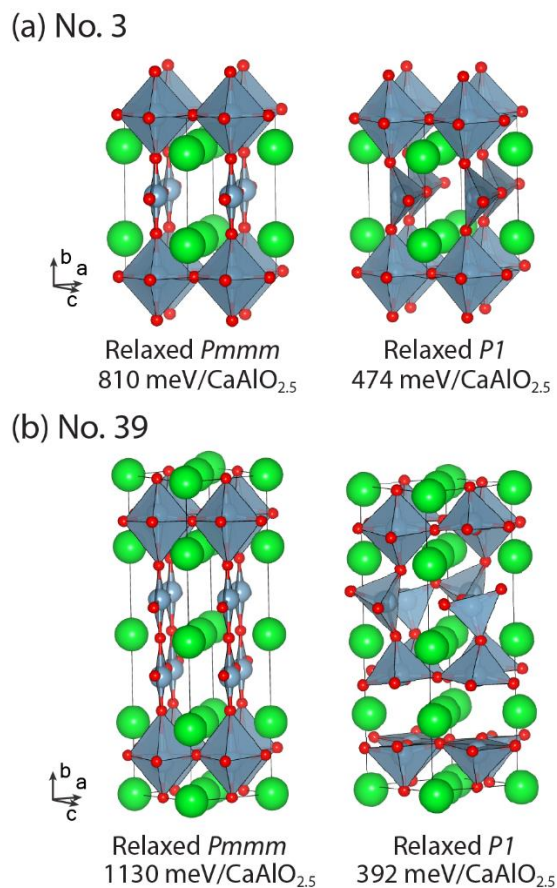

**Figure S7** | Atomic relaxation of CaAlO₂.₅ OOV structures with square planar units: (a) Structure No. 3 and (b) Structure No.39. When symmetry of the structure is broken by a small perturbation to the atomic coordinates, atomic relaxation reconstructs AlO₄ square planar units to tetrahedral or square pyramidal units. This change in coordination is accompanied by a significant decrease in the relative energy difference compared to lowest energy OOV phase. Ca, Al, and O atoms are represented by large green, blue, and small red spheres, respectively.



\end{document}